\documentclass[11pt,a4paper]{article}

\usepackage{jheppub}
\usepackage{chessfss}
\newcommand{\beqa}{\begin{eqnarray}}
\newcommand{\eeqa}{\end{eqnarray}}

\newcommand{\beq}{\begin{equation}}
\newcommand{\eeq}{\end{equation}}
\newcommand{\bea}{\begin{eqnarray}}
\newcommand{\eea}{\end{eqnarray}}

\newcommand{\CF}{{\mathcal F}}

\newcommand{\CN}{{\mathcal N}}

\newcommand{\CT}{{\mathcal T}}

\newcommand\SU{{\rm SU}}
\newcommand\U{{\rm U}}

\def\Tr{\mathop{\rm Tr}}
\newcommand\tr{\mathrm{tr}}

\newcommand\diag{\mathrm{diag}}

\newcommand{\be}{\begin{equation}}
\newcommand{\ee}{\end{equation}}
\newcommand{\cG}{{\cal U}}
\newcommand{\bfa}{{\bf a}}
\newcommand{\bfb}{{\bf b}}
\newcommand{\bfc}{{\bf c}}
\newcommand{\bfd}{{\bf d}}
\newcommand{\bfu}{{\bf u}}
\newcommand{\mn}{\star}

\def\tr{\mathop{\mathrm{tr}}\nolimits}

\def\SU{\mathrm{SU}}
\def\UU{\mathrm{U}}

\def\cN{\mathcal{N}}

\def\vev#1{\langle#1\rangle}

\def\SU{\mathrm{SU}}

\def\cN{\mathcal{N}}
\def\cT{\mathcal{T}}
\def\cU{\mathcal{U}}

\def\diag{\mathop{\mathrm{diag}}}
\def\tr{\mathop{\mathrm{tr}}\nolimits}
\def\vev#1{\langle#1\rangle}

\def\SU{SU}
\def\U{U}
\def\UU{U}

\usepackage{tikz}
\usetikzlibrary{arrows}
\tikzset{>=stealth}

\setboardfontsize{8}

\title{New $\CN{=}1$ Dualities}

\preprint{CALT-68-2917, IPMU-13-0054, UT-13-07}

\author[\WhiteKnightOnWhite]{Abhijit Gadde,}
\affiliation[\WhiteKnightOnWhite]{California Institute of Technology \\ Pasadena, CA 91125, USA}
\emailAdd{abhijit@theory.caltech.edu}

\author[\WhiteKnightOnWhite]{Kazunobu Maruyoshi,}
\emailAdd{maruyosh@caltech.edu}

\author[\BlackBishopOnWhite]{Yuji Tachikawa,}
\affiliation[\BlackBishopOnWhite]{Department of Physics, Faculty of Science, \\
 University of Tokyo,  Bunkyo-ku, Tokyo 133-0022, Japan and \\
Institute for the Physics and Mathematics of the Universe, \\
 University of Tokyo,  Kashiwa, Chiba 277-8583, Japan}
\emailAdd{yuji.tachikawa@ipmu.jp}

\author[\WhiteKnightOnWhite]{and Wenbin Yan}
\emailAdd{wbyan@theory.caltech.edu}

\abstract{
We show that the  $\cN{=}1$ supersymmetric $\SU(N)$ gauge theory with $2N$ flavors without superpotential has not only the standard Seiberg dual description but also another dual description involving two copies of the so-called $T_N$ theory. This is a natural generalization to $N>2$ of a dual description of $\SU(2)$ gauge theory with 4 flavors found by Csaki, Schmaltz, Skiba and Terning. We also study dualities of other $\cN{=}1$ SCFTs involving copies of $T_N$ theories. 
Our duality is the basic operation from which a recently-found web of $\cN{=}1$ dualities obtained by compactifying M5-branes on Riemann surfaces can be derived field-theoretically. 
}


\begin{document}
\setcounter{tocdepth}{2}
\maketitle
%%%%%%%%%%%%%%%% section 1 %%%%%%%%%%%%%%%%%%%%%%%%%%%%%%%%%%%%%%%%
\section{Introduction and summary}
\label{sec:intro}

The main aim of this paper is to present a third dual description of the simplest of supersymmetric dual pairs introduced by Seiberg \cite{Seiberg:1994pq}, namely \begin{itemize}
\item $\cN{=}1$ $\SU(N)$ theory with $2N$ flavors with zero superpotential, and
\item $\cN{=}1$ $\SU(N)$ theory with $2N$ flavors $Q$, $\tilde Q$ and $4N^2$ singlets $M$ with superpotential $W=MQ\tilde Q$.
\end{itemize} 
It has the following non-conventional form, given by 
\begin{itemize}
\item $\cN{=}1$ $\SU(N)$ theory coupled to two copies of the so-called $T_N$ theory first introduced in \cite{Gaiotto:2009we}, together with $2N^2+2N$ singlets and a specific superpotential which is given in \eqref{thirddualsuperpotential}. 
\end{itemize}
This third description is a generalization of a dual description of $\cN{=}1$ $\SU(2)$ gauge theory with four flavors found in \cite{Csaki:1997cu}, which exists in addition to the standard duals \cite{Seiberg:1994pq,Intriligator:1995ne}. This exemplifies a general observation in the last few years that, to extend known supersymmetric dualities among conventional supersymmetric gauge theories in the low-rank gauge groups to bigger gauge groups, one needs to consider non-conventional theories which involve less familiar ingredients such as the $T_N$ theory.

\subsection{Philosophical digression} 
Let us reflect on this general observation, before moving on to the technical discussions.
A busy reader can skip to section~\ref{subsec:mainideas}.

\medskip

A \emph{conventional QFT} is composed of the three ingredients: 
\begin{enumerate}
\item A selection of free matter fields,
\item Gauge fields coupled to flavor symmetry currents of the item 1 above, and
\item Gauge-invariant interaction terms of the item 1 and item 2.
\end{enumerate} 
We now have a considerable set of techniques to analyze its behavior.
Note that a set of free matter fields is just one example of a conformal theory, namely a trivial conformal field theory. Therefore we can consider a larger class of quantum field theory with the following structure, generalizing the conventional ones \cite{Georgi:2007ek,Meade:2008wd}: 
\begin{enumerate}
\item[1'.] A selection of (possibly non-trivial) conformal theories,
\item[2.] Gauge fields coupled to flavor symmetry currents of the item 1', and
\item[3.] Gauge-invariant interaction terms of the item 1' and item 2.
\end{enumerate}
Here, the conformal theory we use as the ingredients in the item 1' does not have to be realized as an infrared or strongly-coupled limit of a conventional QFT. In general a single conformal theory can have many such realizations, and we do not want to specify one. We can also imagine a conformal theory such that it does not have any description as a conventional QFT, but that we know its existence using more transcendental methods such as M-theory. What matters here is that we know the properties of the said conformal theory in sufficient detail to analyze  the combined system in a meaningful way, and the question is if we can actually do this.

With supersymmetry, we are now in a position to perform such analyses. It's not just that we are able but we are forced to consider these non-conventional supersymmetric QFTs, if we are interested in supersymmetric dualities in any way, as the following examples amply show:
\begin{itemize}
\item The $\cN{=}2$ $\SU(2)$ gauge theory with four flavors has a triality with itself. Its natural generalization to gauge groups larger than $\SU(2)$ requires non-conventional theories, e.g.~\cite{Argyres:2007cn,Argyres:2007tq,Gaiotto:2009we,Argyres:2010py,Chacaltana:2010ks}. This example taught us that there are many so-far unknown superconformal field theories (SCFTs), the most important of which is called the $T_N$ theory. 
\item The superconformal indices of $\cN{=}2$ theories can be naturally reformulated in terms of 2d topological QFTs only by including non-conventional theories \cite{Gadde:2011ik,Gadde:2011uv,Gaiotto:2012xa}.
\item The dualities of non-conventional $\cN{=}2$ theories  above can be turned into dualities among  non-conventional $\cN{=}1$ theories by an addition of an $\cN{=}1$ superpotential \cite{Benini:2009mz,Bah:2011je,Bah:2011vv,Bah:2012dg,Beem:2012yn}.
\item The most singular point in the Coulomb branch of $\cN{=}2$ $\SU(2)$ gauge theory with one flavor is a single nontrivial SCFT \cite{Argyres:1995jj,Argyres:1995xn}. The most singular point of a general $\cN{=}2$ gauge theory with a number of flavor hypermultiplets is instead described as a non-conventional QFT, where two almost decoupled conformal sectors are coupled by an infrared free magnetic gauge multiplet \cite{Gaiotto:2010jf,Giacomelli:2012ea}. This structure is important to be consistent with the $a$-theorem \cite{Komargodski:2011vj}.
\item Even the most singular point of $\cN{=}2$ pure gauge theory with some gauge groups is better described as a non-conventional theory, where a nontrivial SCFT whose flavor symmetry is gauged by a $\UU(1)$ multiplet \cite{Argyres:2012fu}.
\item In this paper we present a third dual description of the standard Seiberg dual pair. This is another example which reinforces this general picture described above. 
\end{itemize}

In the papers published in the last few years,  non-conventional theories as defined above were often called non-Lagrangian theories.  Given a theory, it is fundamentally ill-defined whether it has a Lagrangian or not. Therefore the authors think  our terminology would be more appropriate. 

\subsection{Main ideas}
\label{subsec:mainideas}
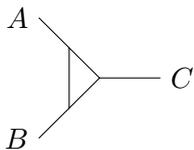
\begin{figure}[ht]
\centering
\begin{tikzpicture}[scale=.4,baseline=(current bounding box.east)]
\draw (0,0)--(-1,-1)--(-1,1)--cycle;
\draw (-2,2) node[anchor=east] {$A$} --(-1,1);
\draw (-2,-2) node[anchor=east] {$B$} --(-1,-1);
\draw (2,0) node[anchor=west] {$C$} --(0,0);
\end{tikzpicture}
\caption{The $T_N$ theory. The vertices labelled by $A,B$ and $C$ represent the flavor symmetries $SU(N)_A$, $SU(N)_B$ and $SU(N)_C$ respectively.}
\label{TNtheory}
\end{figure}

Let us explain our main ideas below.  We begin by recalling the main ingredient: 

\paragraph{The $T_N$ theory}
The  $T_N$ theory is an $\cN{=}2$ superconformal field theory with $\SU(N)_A\times \SU(N)_B\times \SU(N)_C$ flavor symmetry \cite{Gaiotto:2009we}. In particular, it has scalar chiral primary operators $\mu_A$, $\mu_B$, $\mu_C$ of dimension 2, each transforming as an adjoint under one of the three $\SU(N)^3$ symmetries. We denote it graphically as in Fig.~\ref{TNtheory}.

In addition, this theory has  operators $Q_{(k)}$ of scaling dimension $k(N-k)$ transforming in $(\wedge^k, \wedge^k, \wedge^k)$ representation, where $\wedge^k$ is the antisymmetric $k$-index tensor representation. More about these operators will be detailed elsewhere.  The familiar operators $Q_{ijk}$ and $Q^{ijk}$ are simply $Q_{(1)}$ and $Q_{(N-1)}$ respectively, while $Q_{(0)}$ and $Q_{(N)}$ are both identity operators.  
The operators $\mu$ satisfy \cite{Benini:2009mz}
\begin{equation}
\tr \mu_A{}^2 =
\tr \mu_B{}^2 =
\tr \mu_C{}^2\label{relations}.
\end{equation} 

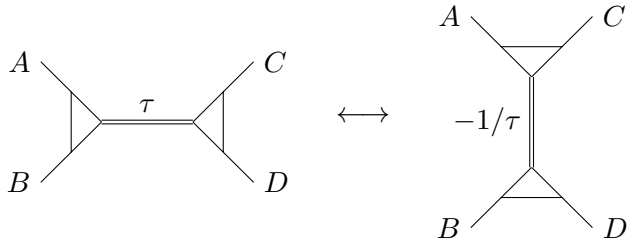
\begin{figure}[ht]
\centering
\[
\begin{tikzpicture}[scale=.4,baseline=(current bounding box.east)]
\draw (0,0)--(-1,-1)--(-1,1)--cycle;
\draw (3,0)--(4,-1)--(4,1)--cycle;
\draw[double,double distance=1pt] (0,0)--(3,0) node[pos=.5,above]{$\tau$};
\draw (-2,2) node[anchor=east] {$A$} --(-1,1);
\draw (-2,-2) node[anchor=east] {$B$} --(-1,-1);
\draw (5,2) node[anchor=west] {$C$} --(4,1);
\draw (5,-2) node[anchor=west] {$D$} --(4,-1);
\end{tikzpicture}
\quad\longleftrightarrow\quad
\begin{tikzpicture}[scale=.4,baseline=(coupling.east)]
\draw (0,0)--(-1,-1)--(1,-1)--cycle;
\draw (0,3)--(-1,4)--(1,4)--cycle;
\draw[double,double distance=1pt] (0,0)--(0,3) node[pos=.5,anchor=east](coupling){$-1/\tau$};
\draw (2,-2) node[anchor=west] {$D$} --(1,-1);
\draw (-2,-2) node[anchor=east] {$B$} --(-1,-1);
\draw (2,5) node[anchor=west] {$C$} --(1,4);
\draw (-2,5) node[anchor=east] {$A$} --(-1,4);
\end{tikzpicture}
\]
\caption{Duality of the theory obtained by coupling two $T_N$ theories with an ${\cal N}=2$ vector multiplet. The double line connecting the $T_N$ blocks stands for the ${\cal N}=2$ vector multiplet.}
\label{N=2duality}
\end{figure}

 The $T_N$ theory behaves particularly regularly under $\cN{=}2$ dualities. For example, consider the theory obtained by gauging the diagonal $\SU(N)$ flavor symmetry of two copies of the $T_N$ theory with ${\cal N}=2$ vector multiplet. The resulting theory at gauge coupling constant $\tau$ is dual to another theory with gauge coupling $-1/\tau$ and reordered flavor symmetry groups. This duality is graphically represented in Fig.~\ref{N=2duality}.

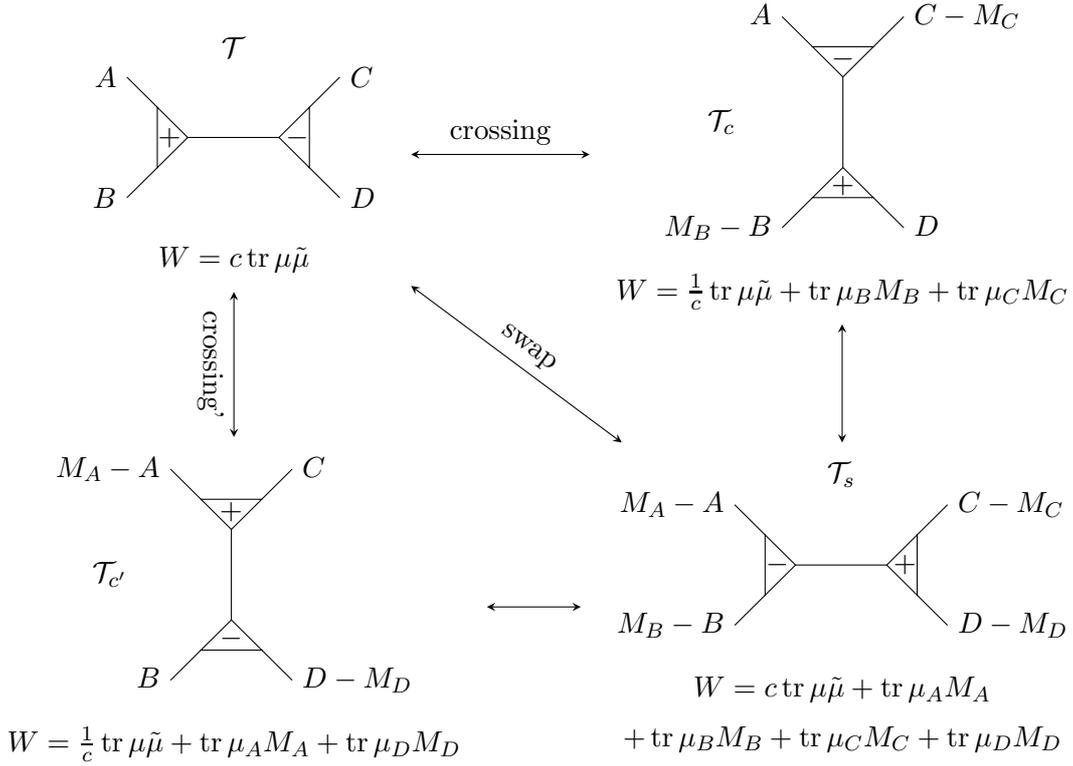
\begin{figure}[th]
\centering\begin{tikzpicture}[scale=5]
\draw (0,1.2) node(o) {\begin{tabular}{c}
	\begin{tikzpicture}[scale=.4]
	\draw (1.5,3) node {	$\mathcal T$ };
	\draw (0,0)--(-1,-1)--(-1,1)--cycle;
	\draw (3,0)--(4,-1)--(4,1)--cycle;
	\draw (0,0)--(3,0) ;
	\draw (-.6,0) node {$+$};
	\draw (3.6,0) node {$-$};
	\draw (-2,2) node[anchor=east] {$A$} --(-1,1);
	\draw (-2,-2) node[anchor=east] {$B$} --(-1,-1);
	\draw (5,2) node[anchor=west] {$C$} --(4,1);
	\draw (5,-2) node[anchor=west] {$D$} --(4,-1);
	\end{tikzpicture} \\
	$W=c\tr \mu \tilde\mu$
\end{tabular}};
\draw (1.6,1.2) node(c) {\begin{tabular}{c}
	\begin{tikzpicture}[scale=.4]
	\draw (-4,1.5) node {	$\mathcal T_c$ };
	\draw (0,0)--(-1,-1)--(1,-1)--cycle;
	\draw (0,3)--(-1,4)--(1,4)--cycle;
	\draw (0,0)--(0,3);
	\draw (0,-.6) node {$+$};
	\draw (0,3.6) node {$-$};
	\draw (2,-2) node[anchor=west] {$D\vphantom{M_D}$} --(1,-1);
	\draw (-2,-2) node[anchor=east] {$M_B-B$} --(-1,-1);
	\draw (2,5) node[anchor=west] {$C-M_C$} --(1,4);
	\draw (-2,5) node[anchor=east] {$A\vphantom{M_A}$} --(-1,4);
	\end{tikzpicture}\\
	$W=\frac1c\tr \mu\tilde\mu + \tr \mu_B M_B+\tr \mu_C M_C $
\end{tabular}};
\draw (0,0) node(cp) {\begin{tabular}{c} 
	\begin{tikzpicture}[scale=.4]
	\draw (-4,1.5) node {	$\mathcal T_{c'}$ };
	\draw (0,0)--(-1,-1)--(1,-1)--cycle;
	\draw (0,3)--(-1,4)--(1,4)--cycle;
	\draw (0,0)--(0,3);
	\draw (0,-.6) node {$-$};
	\draw (0,3.6) node {$+$};
	\draw (2,-2) node[anchor=west] {$D-M_D$} --(1,-1);
	\draw (-2,-2) node[anchor=east] {$B\vphantom{M_B}$} --(-1,-1);
	\draw (2,5) node[anchor=west] {$C\vphantom{M_C}$} --(1,4);
	\draw (-2,5) node[anchor=east] {$M_A-A$} --(-1,4);
	\end{tikzpicture}\\
	$W=\frac1c\tr \mu\tilde\mu + \tr \mu_A M_A+\tr \mu_D M_D $
\end{tabular}};
\draw (1.6,0) node(s) {\begin{tabular}{c}
	\begin{tikzpicture}[scale=.4]
	\draw (1.5,3) node {	$\mathcal T_{s}$ };
	\draw (0,0)--(-1,-1)--(-1,1)--cycle;
	\draw (3,0)--(4,-1)--(4,1)--cycle;
	\draw (0,0)--(3,0);
	\draw (-.6,0) node {$-$};
	\draw (3.6,0) node {$+$};
	\draw (-2,2) node[anchor=east] {$M_A-A$} --(-1,1);
	\draw (-2,-2) node[anchor=east] {$M_B-B$} --(-1,-1);
	\draw (5,2) node[anchor=west] {$C-M_C$} --(4,1);
	\draw (5,-2) node[anchor=west] {$D-M_D$} --(4,-1);
	\end{tikzpicture}\\
	$W=c\tr \mu\tilde\mu + \tr \mu_A M_A $\\
	$+\tr \mu_B M_B+\tr \mu_C M_C +\tr \mu_D M_D$
\end{tabular}};
\draw[<->] (o) -- (c) node[pos=.5,above]{crossing};
\draw[<->] (o) -- (cp) node[pos=.5,sloped,anchor=north]{crossing'};
\draw[<->] (o) -- (s) node[pos=.5,right,sloped,anchor=south]{swap};
\draw[<->] (c)--(s);
\draw[<->] (cp)--(s);
\end{tikzpicture}
\caption{Dualities of the $\cN{=}1$ supersymmetric theory $\cal T$.}
\label{TheoryT}
\end{figure}

\paragraph{$\cal T$ and its dualities}
We now apply the technique  of the inherited duality \cite{Argyres:1996eh,Argyres:1999xu} to the above $\cN{=}2$ dual to derive a duality between two $\cN{=}1$ theories. Instead of the $\cN{=}2$ vector multiplet, the $\cN{=}1$ vector multiplet now gauges the diagonal $\SU(N)$ symmetry of the two $T_N$ theories. The resulting $\cN{=}1$ theory is one of the main focus of this paper. We call this the theory $\cal T$. Inherited from the $\cN{=}2$ theory, $\cal T$ admits new dual descriptions. In this paper, by duality we always mean different UV descriptions which flow to the \emph{same} infrared SCFT i.e.~the same point on the conformal manifold. Different dual descriptions of $\cal T$ are summarized in  Fig.~\ref{TheoryT}.

The meaning of the $\pm$ sign assigned to each $T_N$ theory will become clear in due course. We have also taken this opportunity to label each UV theory. The subscript $c$ in ${\cal T}_c$ and $T_{c^{\prime}}$  stands for \emph{crossing}. The subscript  $s$ in ${\cal T}_s$ stands for \emph{swap}. The symbol $M_X$ denotes the gauge singlet field transforming under the adjoint representation of $\SU(N)_X$. The notation $M_X-X$ represents the superpotential  coupling $\mbox{tr}\, \mu_X M_X$. We have also indicated the superpotential in each theory. We will be more interested in the case when the coefficient $c$ in the superpotential vanishes. In that case, the duality between ${\cal T}$ and ${\cal T}_s$ will interest us the most. Under this duality  the nontrivial operator $\mu_A$ in $\cal T$ is mapped to $M_A$ in ${\cal T}_s$, rather as in the standard Seiberg duality where the nontrivial quadratic mesons on the electric side corresponds to the gauge singlet mesons on the magnetic side. Similar map holds between $\mu_B$ of $\cal T$ and $M_B$ of ${\cal T}_s$, and so on.

\paragraph{$\cG$ and its dualities: Higgsing down to SQCD} 
The $T_N$ theory is the parent of most of the new nontrivial SCFTs found in the last few years, e.g.~in \cite{Chacaltana:2010ks}, in the sense that they are obtained by giving a \emph{nilpotent} vev to some or all of $\mu_{A,B,C}$ and going to the infrared \cite{Benini:2009gi}. This operation is often called the \emph{closure of the puncture} in the literature, since the $T_N$ theory and its likes are the low-energy limit of $N$ M5-branes on a sphere with three punctures of various types. In particular, by setting $\vev{\mu_C}$ to be a matrix $\rho_\mn$ which is nilpotent with Jordan blocks of sizes $1$ and $N-1$ and specifies an embedding of $\SU(2)$ inside $\SU(N)_{C}$, the theory flows to a bifundamental hypermultiplet of $\SU(N)_A\times \SU(N)_B$ with free hypermultiplets \cite{Chacaltana:2012zy}.

We apply this procedure to the punctures $A$ and $D$ in Fig. \ref{TheoryT}. We give vevs $\vev{\mu_A}=\vev{\mu_D}=\rho_\mn$ in ${\cal T}$. The theory $\cal T$ becomes  $\cN{=}1$ $\SU(N)$  gauge theory coupled to $2N$ flavors which we call $\cG$. As $\mu_A$ and $\mu_D$ are mapped to the same $\mu_A$ and $\mu_D$ operators of ${\cal T}_c$, the theory ${\cal T}_c$ also becomes $\cN{=}1$ $\SU(N)$ theory coupled to $2N$ flavors, albeit with adjoint fields $M_B$ and $M_C$. We label this theory as $\cG_c$. This shows that $\cG$ and $\cG_c$ are related by the Seiberg duality. Things are more interesting and involved when we perform the corresponding Higgsing in theories ${\cal T}_s$ and ${\cal T}_{c^\prime}$. Instead of $\mu$, we have to  give vevs $\vev{M_A}=\vev{M_D}=\rho_\mn$. This corresponds to an addition of a superpotential linear in a particular nilponent direction of $\mu_A$ and $\mu_D$. This type of superpotential deformation was studied in \cite{Heckman:2010qv}, which we employ. For example, in the theory ${\cal T}_s$, the $\SU(N)$ adjoint fields $M_A$ and $M_D$ reduce to $2(N+1)$ singlets, $M_{A,i}$ and $M_{D,i}$, 
($i=0, 1, \ldots, N$). The superpotential is
\begin{equation}
W=  \tr \rho_\mn\mu_A + \mu_{A,i} M_{A,i} + \tr \mu_B M_B  
+ \tr \mu_C M_C + \tr \rho_\mn \mu_D + \mu_{D,i} M_{D,i}.   \label{thirddualsuperpotential}
\end{equation}
Here $\mu_{A,i}$ and $\mu_{D,i}$ are the components of $\mu_{A}$ and $\mu_{D}$ which commute with $\rho_{\mn}{}^T$.

\begin{figure}[t]
\centering
\begin{tikzpicture}[scale=5]
\draw (0,1.2) node(o) {\begin{tabular}{c}
	\begin{tikzpicture}[scale=.4]
	\draw (1.5,3) node {	$\mathcal U$ };
	\draw (0,0)--(-1,-1)--(-1,1)--cycle;
	\draw (3,0)--(4,-1)--(4,1)--cycle;
	\draw (0,0)--(3,0) ;
	\draw (-.6,0) node {$+$};
	\draw (3.6,0) node {$-$};
	\fill (-2,2) circle[radius=.2];
	\draw (-2,2) node[anchor=east] {$U(1):\ A$} --(-1,1);
	\draw (-2,-2) node[anchor=east] {$B$} --(-1,-1);
	\draw (5,2) node[anchor=west] {$C$} --(4,1);
	\fill (5,-2) circle[radius=.2];
	\draw (5,-2) node[anchor=west] {$D\ :U(1)$} --(4,-1);
	\end{tikzpicture} \\
	${\mathcal N}{=}1$ $SU(N)$ SQCD with $N_f=2N$
\end{tabular}};
\draw (1.6,1.2) node(c) {\begin{tabular}{c}
	\begin{tikzpicture}[scale=.4]
	\draw (-4,1.5) node {	$\mathcal U_c$ };
	\draw (0,0)--(-1,-1)--(1,-1)--cycle;
	\draw (0,3)--(-1,4)--(1,4)--cycle;
	\draw (0,0)--(0,3);
	\draw (0,-.6) node {$+$};
	\draw (0,3.6) node {$-$};
	\fill (2,-2) circle[radius=.2];
	\draw (2,-2) node[anchor=west] {$D\ :U(1)\vphantom{M_D}$} --(1,-1);
	\draw (-2,-2) node[anchor=east] {$M_B-B$} --(-1,-1);
	\draw (2,5) node[anchor=west] {$C-M_C$} --(1,4);
	\fill (-2,5) circle[radius=.2];
	\draw (-2,5) node[anchor=east] {$U(1):\ A\vphantom{M_A}$} --(-1,4);
	\end{tikzpicture}\\
	${\mathcal N}{=}1$ $SU(N)$ SQCD with $N_f=2N$
\end{tabular}};
\draw (0,0) node(cp) {\begin{tabular}{c} 
	\begin{tikzpicture}[scale=.4]
	\draw (-4,1.5) node {	$\mathcal U_{c'}$ };
	\draw (0,0)--(-1,-1)--(1,-1)--cycle;
	\draw (0,3)--(-1,4)--(1,4)--cycle;
	\draw (0,0)--(0,3);
	\draw (0,-.6) node {$-$};
	\draw (0,3.6) node {$+$};
	\draw (2,-2) node[anchor=west] {$D-M_{D,i}$} --(1,-1);
	\draw (-2,-2) node[anchor=east] {$B\vphantom{M_{B,i}}$} --(-1,-1);
	\draw (2,5) node[anchor=west] {$C\vphantom{M_{D,i}}$} --(1,4);
	\draw (-2,5) node[anchor=east] {$M_{A,i}-A$} --(-1,4);
	\end{tikzpicture}
\end{tabular}};
\draw (1.6,0) node(s) {\begin{tabular}{c}
	\begin{tikzpicture}[scale=.4]
	\draw (1.5,3) node {	$\mathcal U_{s}$ };
	\draw (0,0)--(-1,-1)--(-1,1)--cycle;
	\draw (3,0)--(4,-1)--(4,1)--cycle;
	\draw (0,0)--(3,0);
	\draw (-.6,0) node {$-$};
	\draw (3.6,0) node {$+$};
	\draw (-2,2) node[anchor=east] {$M_{A,i}-A$} --(-1,1);
	\draw (-2,-2) node[anchor=east] {$M_B-B$} --(-1,-1);
	\draw (5,2) node[anchor=west] {$C-M_C$} --(4,1);
	\draw (5,-2) node[anchor=west] {$D-M_{D,i}$} --(4,-1);
	\end{tikzpicture}
\end{tabular}};
\draw[<->] (o) -- (c) node[pos=.5,above]{crossing};
\draw[<->] (o) -- (cp) node[pos=.5,sloped,anchor=north]{crossing'};
\draw[<->] (o) -- (s) node[pos=.5,right,sloped,anchor=south]{swap};
\draw[<->] (c)--(s);
\draw[<->] (cp)--(s);
\end{tikzpicture}
\caption{Different dual descriptions of $\cN{=}1$ $SU(N)$ SQCD with $N_f=2N$. This theory is called $\cG$ in the paper. The black dot represents Higgsing of the $SU(N)$ flavor symmetry associated to at that puncture down to $U(1)$. Theory $\cG_C$ is the conventional Seiberg dual theory while the theories $\cG_{c^\prime}$ and $\cG_s$ are new duals of the SQCD.}
\label{TheoryU}
\end{figure}
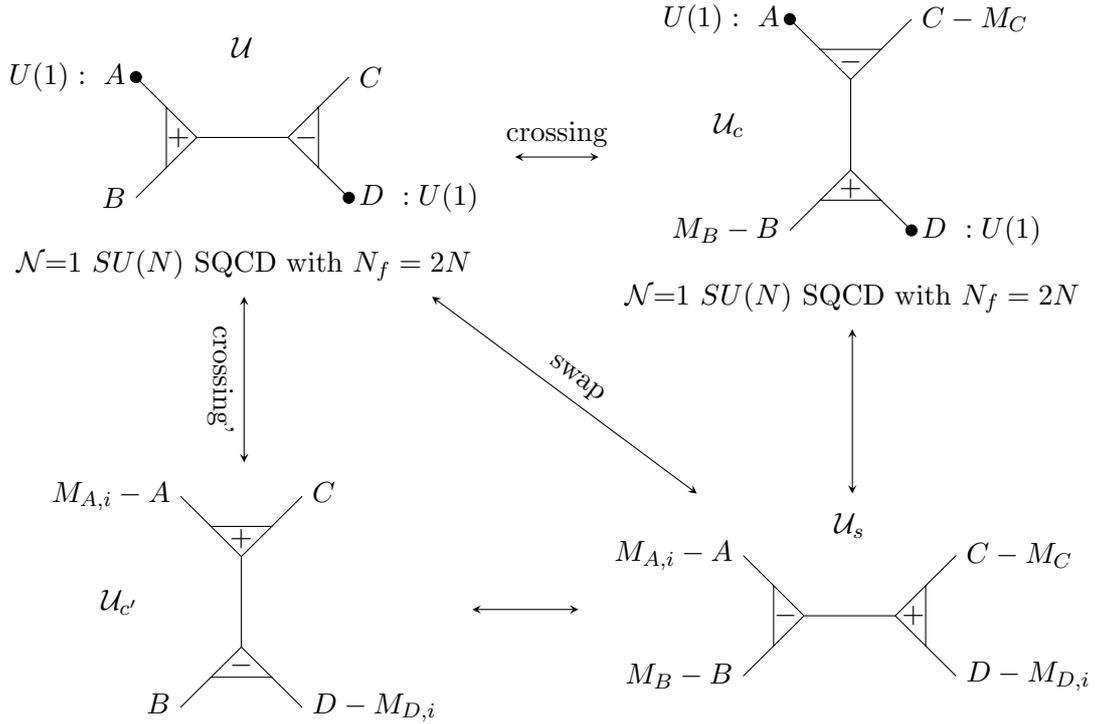

 We call this new dual $\cG_s$. The Higgsing of ${\cal T}_{c^{\prime}}$ can be analyzed in similar fashion and is called $\cG_{c^\prime}$. The dualities of $\cG$ are summarized in Fig.~\ref{TheoryU}.
The duality webs Fig.~\ref{TheoryT} and Fig.~\ref{TheoryU}, in particular the new dualities ${\cal T}\leftrightarrow{\cal T}_s$ and $\cG \leftrightarrow \cG_s$ are put to test by studying the 't Hooft anomaly coefficients  and the superconformal indices. We find nontrivial agreements. We use the previous results about 't Hooft anomalies and indices of the $T_N$ theory from \cite{Gaiotto:2009we,Gaiotto:2009gz,Gadde:2011uv}. 

Note that, instead of $A$ and $D$, we could have Higgsed the punctures $A$ and $C$. This would have led us into another class of theories related to each other by Argyres-Seiberg like dualities. We do not explore this class in the current paper.

\paragraph{With more general punctures} 

We can give a more general nilpotent vev to the operator $\mu_A$ of ${\cal T}$ and the gauge singlet $M_A$ of ${\cal T}_s$. This procedure gives rise to a duality between 
\begin{itemize}
\item $\cN{=}1$ $\SU(N)$ gauge theory coupled to SCFTs obtained by $N$ M5-branes on a sphere with three punctures of general types, and 
\item $\cN{=}1$ $\SU(N)$ gauge theory coupled to two copies of the $T_N$ theory and a number of gauge singlet chiral multiplets with an intricate superpotential generalizing \eqref{thirddualsuperpotential}.
\end{itemize}
We also perform the checks  of the anomaly coefficients and the indices in this generalized case. These anomalies and indices are computed by using their values for SCFTs corresponding to general three punctured spheres  as proposed in \cite{Gadde:2011uv,Chacaltana:2012zy,Gaiotto:2012xa}. In a sense our computation gives an independent confirmation of the proposals made there, as we only use the knowledge of the basic duality ${\cal T}\leftrightarrow{\cal T}_s$ and the fact that the puncture can be closed via giving nilpotent vevs. For example, the index of a general three-punctured sphere has been conjectured to be given roughly by the form 
\begin{equation}
K_{\Lambda_1}(a)K_{\Lambda_2}(b)K_{\Lambda_3}(c)
\sum_\lambda \Psi_\lambda(at^{\Lambda_1})\Psi_\lambda(bt^{\Lambda_2})\Psi_\lambda(ct^{\Lambda_3}),
\end{equation} 
where $\Lambda_{1,2,3}$ denote the type of the punctures, $a,b,c$ the fugacities of flavor symmetries of the punctures, $\Psi_\lambda$ are certain orthogonal polynomials of type $A_{N-1}$ labeled by its representations, and $K_\Lambda(a)$ are prefactors depending on the type of punctures. Our analysis in the paper makes clear that the prefactor $K_\Lambda(a)$ counts the contribution from the conserved current multiplets of the flavor symmetry together with the semi-conserved multiplets, both of which are associated to the puncture of type $\Lambda$, although this fact was used in \cite{Gaiotto:2012uq} to determine the index contribution of non-maximal punctures in class $\cal S$ ${\cal N}=2$ theories.
In this paper, we denote the vev corresponding to the $U(1)$ puncture and no puncture as $\rho_\mn$ and $\rho_\varnothing$ respectively.

\paragraph{Quivers and relation to the geometry of M5-branes}

Equipped with the dualities Fig.~\ref{TheoryT} and its Higgsed versions Fig.~\ref{TheoryU}, we can study $\cN{=}1$ supersymmetric generalized quivers with $T_N$ theories. As will be shown in section~\ref{sec:quiver}, $T_N$ theories with opposite sign assignments are coupled with the ${\cal N}=1$ vector multiplet and $T_N$ theories with the same sign are coupled with the $\cN{=}2$ vector multiplet. Similar quiver construction involving $T_N$ theories arose in \cite{Benini:2009mz,Bah:2011je,Bah:2011vv,Bah:2012dg,Beem:2012yn} while identifying the UV description of the $\cN{=}1$ SCFTs obtained from M5-brane compactifications. In \cite{Bah:2011vv,Bah:2012dg}, authors only studied $\cN{=}1$ SCFTs with no flavor symmetry i.e.~the theories obtained from compactifying M5-branes on Riemann surfaces without puncture. The superconformal index of the theories corresponding to the Riemann surface with maximal punctures was computed in \cite{Beem:2012yn}.
The generalized quivers studied in this work do have flavor symmetries and hence correspond to $\cN{=}1$ compactifications on surfaces with (not necessarily maximal) punctures. 
We will see that adjoint $M_X$ fields appearing in our duality can be thought of as arising from the scalar component of $\cN{=}2$ vector multiplet coupled to the puncture.

\subsection{Organization} 
The rest of the paper is organized as follows. In Section \ref{sec:fieldtheory}, we propose and derive the new dual description $\cG_s$ of $\cN{=}1$ $\SU(N)$ gauge theory with $2N$ flavors. In the process we also derive the dualities of ${\cal T}$. We perform checks of the conjectured dualities by computing 't Hooft anomaly coefficients in Section \ref{sec:anomaly} and the superconformal indices in Section \ref{sec:index}. In Section \ref{sec:quiver}, we consider generalized quivers of $T_N$ theories with $\cN{=}1$ supersymmetry and study the relation of this class of theories to the $\cN{=}1$ theories of class $\cal S$. 

%%%%%%%%%%%%%%%% section 2 %%%%%%%%%%%%%%%%%%%%%%%%%%%%%%%%%%%%%%%%%%
\section{Motivation and derivation}
\label{sec:fieldtheory}
In this section, we present our duality proposal. It is motivated in subsection \ref{csaki}.  In subsection \ref{subsec:dualTN} 
we first introduce the \emph{crossing} duality ${\cal T}\leftrightarrow{\cal T}_c$ which squares to our main interest, the \emph{swap} duality ${\cal T}\leftrightarrow {\cal T}_s$. 
We show how the crossing duality is inherited from the S-duality  \cite{Gaiotto:2009we} of the parent $\cN{=}2$ theory. In subsection \ref{subsec:close} we consider the Higgsed version of the duality: $\cG \leftrightarrow \cG_s$. We explore in detail $\cG_s$, the new dual of SQCD.

%%%%%%%%%%%%%%%%%%%%%%%%%%%%%%%%%%
\subsection{Known facts}\label{csaki}
Let us start the study of the dualities of the theory ${\cal T}$ from $N=2$. The $T_2$ theory is simply the theory of $8$ free chiral multiplets. In this case, $\cal T$ is the familiar $\SU(2)$ SQCD with $N_f=4$. The crossing duality ${\cal T}\leftrightarrow {\cal T}_c$ is the Seiberg duality while the swap ${\cal T}\leftrightarrow{\cal T}_s$ is the duality discovered by Csaki, Schmaltz, Skiba and Terning \cite{Csaki:1997cu}. With this in mind, our swap duality can be considered a generalization of the Csaki et al.~duality. 

Consider $\cN{=}1$ supersymmetric $\SU(2)$ gauge theory with four quarks and four anti-quarks. This theory flows to a non-trivial SCFT in the infrared. The global symmetry group is $\SU(8)\times \U(1)_R$, where $\U(1)_R$ is the R-symmetry.  In order to think of this theory as $\cal T$ for $N=2$, we have to focus on the $\SU(2)^4$ subgroup of the flavor symmetry. Under this subgroup the field content transforms as:
\begin{center}
\begin{tabular}{|c|c|c|c|c|c|c|c|}
\hline 
 & $\SU(2)_g$ & $\SU(2)_{A}$ & $\SU(2)_{B}$ & $\SU(2)_{C}$ & $\SU(2)_{D}$ & $\U(1)_{{\cal F}}$ & $\U(1)_{R}$\tabularnewline
\hline 
\hline 
$q$ & $\square$ & $\square$ & $\square$ & $\cdot$ & $\cdot$ & $-1$ & $\frac{1}{2}$\tabularnewline
\hline 
$\tilde{q}$$ $ & $\square$ & $\cdot$ & $\cdot$ & $\square$ & $\square$ & $+1$ & $\frac{1}{2}$\tabularnewline
\hline 
\end{tabular}
\par\end{center}
Here $\SU(2)_g$ denotes the gauge group.
The theory admits $72$ duality frames in total \cite{Spiridonov:2008zr, Dimofte:2012pd}. They fall in three classes: $35$ Seiberg duals \cite{Seiberg:1994pq} where $q\tilde q$ mesons are \emph{flipped} to gauge singlet fields in the dual theory, $35$ Csaki et al.~duals where the baryons $qq$ and anti-baryons $\tilde q \tilde q$  are flipped and finally a single Intriligator-Pouliot dual where both mesons and baryons are flipped. The multiplicity of Seiberg and Csaki et al.~duals comes from different ways of splitting $8$ chiral multiplets into quarks and anti-quarks. We focus on the Csaki et. al. duality, whose field content is summarized below.
\begin{center}
\begin{tabular}{|c|c|c|c|c|c|c|c|}
\hline 
 & $SU(2)_g$ & $SU(2)_{A}$ & $SU(2)_{B}$ & $SU(2)_{C}$ & $SU(2)_{D}$ & $U(1)_{{\cal F}}$ & $U(1)_{R}$\tabularnewline
\hline 
\hline 
$Q$ & $\square$ & $\square$ & $\square$ & $\cdot$ & $\cdot$ & $+1$ & $\frac{1}{2}$\tabularnewline
\hline 
$\tilde{Q}$ & $\square$ & $\cdot$ & $\cdot$ & $\square$ & $\square$ & $-1$ & $\frac{1}{2}$\tabularnewline
\hline 
\hline 
$M_{A}$ & $\cdot$ & $\mbox{adj}$ & $\cdot$ & $\cdot$ & $\cdot$ & $-2$ & $1$\tabularnewline
\hline 
$M_{B}$ & $\cdot$ & $\cdot$ & $\mbox{adj}$ & $\cdot$ & $\cdot$ & $-2$ & $1$\tabularnewline
\hline 
$M_{C}$ & $\cdot$ & $\cdot$ & $\cdot$ & $\mbox{adj}$ & $\cdot$ & $+2$ & $1$\tabularnewline
\hline 
$M_{D}$ & $\cdot$ & $\cdot$ & $\cdot$ & $\cdot$ & $\mbox{adj}$ & $+2$ & $1$\tabularnewline
\hline 
\end{tabular}
\par\end{center}
The gauge singlet fields $M_X$ ($X=A,B,C,D$) are coupled to the dual quark bilinears via the superpotential
\begin{equation}\label{SU2W}
W= M_A (QQ)_A + M_B (QQ)_B + M_C (\tilde Q\tilde Q)_C + M_D (\tilde Q \tilde Q)_D.
\end{equation}
In this equation, by $(QQ)_X$ or $(\tilde Q\tilde Q)_X$  we have denoted the bilinear transforming in the adjoint representation of the $\SU(2)_X$ flavor symmetry.

\subsection{Dualities of coupled $T_N$ theories}\label{subsec:dualTN}
\subsubsection{The theory $\cT$ }
Let us take two copies of $T_N$ theory, which we distinguish by calling them $T_N$ and $\tilde T_N$.
Pick one $\SU(N)$ flavor symmetry each from $T_N$ and $\tilde T_N$, and couple an $\cN{=}1$ $\SU(N)$ vector multiplet to them.
This is the theory $\cT$. 
The system flows to an SCFT in the infrared \cite{Benini:2009mz}.
The IR $\U(1)_R$ charges 
can be obtained from the $\U(1)_R$ and $\SU(2)_R$ of the parent $\cN{=}2$ theory as \cite{Tachikawa:2009tt}:
\be
R=\frac12 R_{\cN{=}2} + I_3,
\label{chargerelation1}
\ee
where $R$, $R_{\CN{=}2}$ and $I_{3}$ are the generators of the IR $\U(1)_{R}$, $\CN{=}2$ $\U(1)_{R}$ and $\U(1) (\subset \SU(2)_{R})$ symmetries respectively.

As the $T_N$ theory is $\cN{=}2$ supersymmetric, another linear combination of its $\cN{=}2$ R-symmetries is a flavor symmetry with respect to a given $\cN{=}1$ subalgebra, which we denote by $J$:
\be
J=R_{\cN{=}2} - 2 I_{3}.
\label{chargerelation2}
\ee 
Let us denote by $\tilde J$ the corresponding charge of the second copy $\tilde T_N$. Then the anomaly free global symmetry $\U(1)_\CF$ of the coupled theory $\cT$ is
\be
\CF=J-\tilde J.
\ee

The role of quark bilinears $qq$ for $N=2$ is played by the chiral operators $\mu_X$ for $N>2$. They transform in the adjoint representation of the flavor symmetry $\SU(N)_X$. Then the \emph{matter content} of the theory $\cal T$ can be summarized as shown:
\begin{center}
\begin{tabular}{|c|c|c|c|c|c|c|c|}
\hline 
 & $\SU(N)_g$ & $\SU(N)_{A}$ & $\SU(N)_{B}$ & $\SU(N)_{C}$ & $\SU(N)_{D}$ & $\U(1)_{{\cal F}}$ & $\U(1)_{R}$\tabularnewline
\hline 
\hline 
$\mu_{A}$ & $\cdot$ & $\mbox{adj}$ & $\cdot$ & $\cdot$ & $\cdot$ & $-2$ & $1$\tabularnewline
\hline 
$\mu_{B}$ & $\cdot$ & $\cdot$ & $\mbox{adj}$ & $\cdot$ & $\cdot$ & $-2$ & $1$\tabularnewline
\hline 
$\mu_{C}$ & $\cdot$ & $\cdot$ & $\cdot$ & $\mbox{adj}$ & $\cdot$ & $+2$ & $1$\tabularnewline
\hline 
$\mu_{D}$ & $\cdot$ & $\cdot$ & $\cdot$ & $\cdot$ & $\mbox{adj}$ & $+2$ & $1$\tabularnewline
\hline 
\end{tabular}
\par\end{center}
Here $\SU(N)_g$ denotes the gauge group.
The operators $\mu_A,\, \mu_B$ and $\mu_C,\, \mu_D$ come from the $T_N$ and ${\tilde T}_N$ theories respectively. They have opposite $\U(1)_{\CF}$ charges. Here we introduce the notation, where we associate a sign $\pm$ to a $T_N$ theory if its $\mu$ operator has $\U(1)_{\CF}$ charge $\mp 2$. This notation was used in Fig.~\ref{TheoryT}.

\subsubsection{Duality ${\cal T}\leftrightarrow {\cal T}_s$:}

Motivated from the Csaki et al.~duality discussed in the previous subsection, we propose the dual description ${\cal T}_s$: theory $\cal T$ with opposite $\U(1)_\CF$ charge coupled to the gauge singlet field $M_X$ through the superpotential
\be
W=\tr\hat{\mu}_A M_A+
\tr\hat{\mu}_B M_B+ 
\tr\hat{\mu}_C M_C+ 
\tr\hat{\mu}_D M_D.
\ee
The gauge singlet $M_X$ transforms in the adjoint representation of the $SU(N)_X$ flavor symmetry. 
The charges of $\hat{\mu}$ operators and the gauge singlets $M$ on the dual side are summarized below.
\begin{center}
\begin{tabular}{|c|c|c|c|c|c|c|c|}
\hline 
 & $\SU(N)_g$ & $\SU(N)_{A}$ & $\SU(N)_{B}$ & $\SU(N)_{C}$ & $\SU(N)_{D}$ & $\U(1)_{{\cal F}}$ & $\U(1)_{R}$\tabularnewline
\hline 
\hline 
$\hat{\mu}_{A}$ & $\cdot$ & $\mbox{adj}$ & $\cdot$ & $\cdot$ & $\cdot$ & $+2$ & $1$\tabularnewline
\hline 
$\hat{\mu}_{B}$ & $\cdot$ & $\cdot$ & $\mbox{adj}$ & $\cdot$ & $\cdot$ & $+2$ & $1$\tabularnewline
\hline 
$\hat{\mu}_{C}$ & $\cdot$ & $\cdot$ & $\cdot$ & $\mbox{adj}$ & $\cdot$ & $-2$ & $1$\tabularnewline
\hline 
$\hat{\mu}_{D}$ & $\cdot$ & $\cdot$ & $\cdot$ & $\cdot$ & $\mbox{adj}$ & $-2$ & $1$\tabularnewline
\hline 
\hline 
$M_{A}$ & $\cdot$ & $\mbox{adj}$ & $\cdot$ & $\cdot$ & $\cdot$ & $-2$ & $1$\tabularnewline
\hline 
$M_{B}$ & $\cdot$ & $\cdot$ & $\mbox{adj}$ & $\cdot$ & $\cdot$ & $-2$ & $1$\tabularnewline
\hline 
$M_{C}$ & $\cdot$ & $\cdot$ & $\cdot$ & $\mbox{adj}$ & $\cdot$ & $+2$ & $1$\tabularnewline
\hline 
$M_{D}$ & $\cdot$ & $\cdot$ & $\cdot$ & $\cdot$ & $\mbox{adj}$ & $+2$ & $1$\tabularnewline
\hline 
\end{tabular}
\par\end{center}
As we swapped the $\pm$ sign assignments of $T_N$ and $\tilde T_N$,  we call this duality the \emph{swap}.

\subsubsection{Matching of chiral operators}
As in the standard Seiberg duality, the operator $\mu_X$ in the original theory $\cT$ is mapped to gauge singlet chiral matter fields $M_X$ in the dual $\cT_s$. 
Let us sketch the matching of the baryon-like operators.
The $T_N$ theory has operators $Q_{(k)}$ transforming in  $(\wedge^{k} ,\wedge^{k},\wedge^{k})$ of the $SU(N)^3$ flavor symmetry,
where $\wedge^k$ is the $k$-index antisymmetric tensor representation. 
Denote by $\tilde{Q}_{(k)}$ the corresponding operator of another copy $\tilde{T}_{N}$.
 ${\hat Q}_{(k)}$ and $\tilde{\hat{Q}}_{(k)}$ be the operators in $\CT_{s}$, respectively.
Note that $Q_{(k)}$ and $\tilde{\hat{Q}}_{(k)}$ ($\tilde{Q}_{(k)}$ and ${\hat Q}_{(k)}$) have negative (positive) $\CF$-charges. 
Then, the gauge-invariant operators 
\be
B^{(k)}=Q_{(N-k)}{\tilde Q}_{(k)}
\ee
can be constructed in $\CT$. They all have zero $\CF$ charge. 
The dual operator is simply \begin{equation}
B^{(k)}=\hat Q_{(N-k)}{\tilde{\hat Q}}_{(k)}.
\end{equation} 

We will perform further checks of the matching of all the supersymmetric operators in section \ref{sec:index} by comparing the superconformal indices on both sides of the duality. 

%%%%%%%%%%%%%%%%%%
\paragraph{Duality ${\cal T}\leftrightarrow{\cal T}_c$:}
 Consider the theory $\cal T$ with the superpotential 
\be
W=c \,\tr \, \mu \tilde\mu.
\ee
where $\mu$ and $\tilde \mu$ are the operators of the $T_{N}$ and $\tilde{T}_{N}$ theories, transforming in the adjoint representation of the gauge symmetry as we mentioned above. On the other side of the duality we have ${\cal T}_c$ graphically represented in Fig.~\ref{TheoryT}. In addition to having a copy of $\cal T$, it also has gauge singlet fields $M_B$ and $M_C$ with the superpotential
\be
W=\tr {\hat\mu}_B M_B+ \tr {\hat\mu}_C M_C+ \hat c\,\tr\, {\hat\mu} \tilde {\hat\mu}
\ee
where $\hat c\sim 1/c$. 
The operators $\hat \mu$ and $\tilde{\hat \mu}$ come from ${\hat T}_N$ and ${\tilde{\hat T}}_N$ in the dual description. The operators $\mu_B$ and $\mu_C$ of the original theory have been mapped to $M_B$ and $M_C$. On the other hand, $\mu_{A}$ and $\mu_{D}$ are mapped to $\hat{\mu}_{A}$ and $\hat{\mu}_{D}$. 

If we perform the $\text{crossing}^{\prime}$ duality on ${\cal T}_c$, ${\hat\mu}_B$ and ${\hat\mu}_C$ are again mapped to new gauge singlet fields ${\hat M}_B$ and ${\hat M}_C$ and the superpotential is
\be
W=\tr {\hat M}_B M_B+ \tr {\hat M}_C M_C+ c \,\tr\, {\hat{\hat\mu}} \tilde {\hat{\hat\mu}}.
\ee
After integrating out $M_B,M_C$ and ${\hat M}_B,{\hat M}_C$, we get back the same theory. We reach the conclusion, $\text{crossing}\cdot\text{crossing}^{\prime}=1$. 

Instead of the $\text{crossing}^{\prime}$, if we apply the $\text{crossing}$ duality to ${\cal T}_c$, the operators ${\hat\mu}_A$ and $\hat\mu_D$ are mapped to new gauge singlet fields $M_A$ and $M_D$. The superpotential becomes
\be
W=\tr {\hat{\hat \mu}}_A M_A+\tr {\hat{\hat \mu}}_B M_B+ \tr {\hat{\hat \mu}}_C M_C+\tr {\hat{\hat \mu}}_D M_D+ c \,\tr\, {\hat{\hat\mu}} \tilde {\hat{\hat\mu}},
\ee
which is  the superpotential of ${\cal T}_s$. Therefore, we see that $\text{crossing}\cdot \text{crossing}=\text{swap}$. We now proceed to derive the crossing duality, which we have seen to be the generating duality for the web Fig.~\ref{TheoryT}, from the S-duality.

%%%%%%%%%%%%%%%%%%
\subsubsection{Derivation of the crossing from S-duality}
The S-duality of the $\cN{=}2$ gauge theory coupled to two copies of $T_N$ theory was graphically represented in Fig.~\ref{N=2duality}. We break the supersymmetry to $\cN{=}1$ by adding a mass term for the adjoint chiral field $\frac m2 \Phi^2$ to the superpotential. The superpotential is now $W = \tr \Phi (\mu - \tilde{\mu}) + \frac{m}{2} \tr \Phi^{2}$ where the first term is inherited from the $\cN{=}2$ theory. Integrating out the massive field $\Phi$, we get
\be\label{Wo}
W=- \frac{c}{2} \tr \mu^{2} - \frac{c}{2} \tr \tilde{\mu}^{2} + c ~ \tr \mu \tilde{\mu}.
\ee
where $c=\frac1m$. The operators $\mu$ and $\tilde \mu$ have $\U(1)_{\CF}$ charge $-2$ and $+2$, hence the first two terms break the $\U(1)_{\CF}$ symmetry. Using the operator relation $\tr \mu_{A}^{2} = \tr \mu_{B}^{2} = \tr \mu_{C}^{2}$ of the $T_{N}$ theory, we rewrite these two terms as $- \frac{c}{2} \tr \mu^{2}_{B} - \frac{c}{2} \tr \mu^{2}_{C}$, where $\mu_{B}$ and $\mu_{C}$ are in the adjoint representations of the $\SU(N)_{B}$ and $\SU(N)_{C}$ flavor symmetries respectively. These and the last term in \eqref{Wo} are independent exactly marginal operators. By turning off the coupling of the exactly marginal deformation which breaks the $\U(1)_{\CF}$ symmetry, we get
\be
W= c ~ \tr \mu \tilde{\mu}.
\ee 

After integrating out the massive adjoint on the S-dual side, we obtain the similar superpotential as \eqref{Wo}: $W= - \frac{\hat{c}}{2} \tr \hat{\mu}^{2} - \frac{\hat{c}}{2} \tr {\tilde{\hat{\mu}}}^{2} + \hat{c}~ \tr \hat{\mu} {\tilde{\hat{\mu}}}$. The constant $\hat c$ is different from $c$ as the gauge coupling of the dual side is inverse of the original gauge coupling. We expect $\hat c\simeq 1/c$.

Using the operator relation on the dual side, we rewrite the superpotential as
\bea
W
&=&  - \frac{\hat{c}}{2} \tr \hat{\mu}^{2}_{B} - \frac{\hat{c}}{2} \tr \hat{\mu}^{2}_{C} + \hat{c} ~ \tr \hat{\mu} \tilde{\hat{\mu}} \nonumber \\
&=&    \frac{1}{2 \hat{c}} \tr M_{B}^{2} + \tr M_{B} \hat{\mu}_{B} + \frac{1}{2 \hat{c}} \tr M_{C}^{2}+ \tr M_{C} \hat{\mu}_{C} + \hat{c} ~ \tr \hat{\mu} \tilde{\hat{\mu}}.
\label{Wm}
\eea
In the second line, we have \emph{integrated in} the gauge singlet fields $M_B$ and $M_C$. In this superpotential, the first and the third terms break the $U(1)_{\CF}$ symmetry. We tune their exactly marginal coefficient so that the $U(1)_{\CF}$ symmetry is restored as in the original theory $\cal T$. Now the superpotential of the dual theory is simply
\be\label{Wc}
W=\tr M_{B} \hat{\mu}_{B} + \tr M_{C} \hat{\mu}_{C} + \hat{c} ~ \tr \hat{\mu} \tilde{\hat{\mu}}.
\ee
The operators $\mu_B$ and $\mu_C$ of $U(1)_{\CF}$ charges $-2$ and $+2$ are mapped to the gauge singlet fields $M_B$ and $M_C$ on the dual side. Because the superpotential should be neutral under the $U(1)_{\CF}$ symmetry, the charge of dual $\hat{\mu}_{B}$ and $\hat{\mu}_{C}$ is $+2$ and $-2$ respectively. However, the $U(1)_{\CF}$ charges of $\mu_A$ and $\mu_D$ have not been affected on the dual side. From the superpotential \eqref{Wc} and the assignment $U(1)_\CF$ charges we conclude that the dual theory is indeed ${\cal T}_c$. Performing the crossing duality twice gives back the original theory, so the constant ${\hat c}=f(c)$ where $f$ is a function with the property $f(f(c))=c$. Our expectation $f(c) \simeq 1/c$ does obey this relation.

%%%%%%%%%%%%%%%%%%%%%%%%%%%%%%%%%%%%%%%%%%%%%%%
\subsection{Dualities of SQCD}
\label{subsec:close}
In this subsection, we consider the dualities Fig.~\ref{TheoryU} of the theory $\cG$ obtained by Higgsing, or equivalently partially closing the punctures, of the theory ${\cal T}$. Before analyzing the dualities, let us review Higgsing of the $T_N$ theory.

\subsubsection{Review of the Higgsing of the $T_N$ theory}
A puncture of the $A_{N-1}$ theory of class $\cal S$ is classified by the $SU(2)$ embedding
\be
\rho:SU(2)\to SU(N).
\ee
This embedding is described  by specifying how the fundamental representation of $SU(N)$ decomposes into irreducible representations of the embedded $SU(2)$: $\square \to {\bf n_1}+\ldots+ {\bf n_\ell}$. This associates a partition $\Lambda=(n_1,\ldots,n_\ell)$ to each puncture. This also means that $\rho(\sigma^+)$ has a Jordan block structure such that $\alpha$-th block has the size $n_\alpha \times n_\alpha$. The commutant of the embedding is the flavor symmetry $S[\prod_i U(r_i)]$ where
$r_i=\Lambda^T{}_i-\Lambda^T{}_{i+1}$ where $\Lambda^{T}$ is the partition dual to $\Lambda$.
The adjoint representation of $SU(N)$ decomposes as
\bea
\text{adj}=\left[\bigoplus_{\alpha=1}^{\ell} \bigoplus_{j=1}^{n_{\alpha} -1} V_{j} \right]
\oplus (\ell-1) V_{0}\oplus 
2 \left[\bigoplus_{\alpha < \beta} \bigoplus_{k=1}^{n_{\beta}} V_{ \frac{n_{\beta} + n_{\alpha} - 2k}{2} }\right]
\equiv\bigoplus_j R_{j} \otimes V_{j},
\label{dec}
\eea
where $V_j$ is the spin $j$ representation of $SU(2)$ and $R_j$ are the flavor symmetry representations. The first two and the last terms come from the diagonal and off-diagonal blocks.

The $T_N$ theory has three maximal punctures, for which $\rho=0$.
Given a nontrivial $\rho_\Lambda$  corresponding to the partition $\Lambda$, we give the vev
\be
\vev{\mu}=\rho_{\Lambda}(\sigma^+)
\ee
which changes the maximal puncture into a puncture of type $\Lambda$.
This procedure produces $N_{\Lambda}$ neutral free half-hypermultiplets in addition in the process. 
Here, $N_\Lambda$ is the dimension of the orbit of the nilpotent element $\rho_{\Lambda}(\sigma^+)$ and is given by 
\be
N_{\Lambda}=\sum_j 2j\, {\rm dim} R_j=\sum_{\alpha=1}^{\ell}\sum_{j=1}^{n_\alpha-1}2j+\sum_{\alpha<\beta}\sum_{k=1}^{n_\beta}2(n_\beta+n_\alpha-2k).
\ee 

We mostly focus on the puncture $\mn:=(N-1,1)$ with $U(1)$ flavor symmetry. 
After  Higgsing by $\vev{\mu_A}=\rho_\mn(\sigma^+)$, the $T_N$ theory flows to the theory of bi-fundamental hypermultiplets together with $N_{\mn}$ free half-hypermultiplets.
Implicitly, we also use the fact that a maximal puncture can be completely closed with no remnant flavor symmetry by Higgsing it with the vev corresponding to the partition $(N)=:\varnothing$.

%%%%%%%%%%%%%%%
\subsubsection{Seiberg duality: $\cG\leftrightarrow\cG_c$}\label{sd}

Now we consider Higgsing the punctures $A$ and $D$ on both sides of the duality ${\cal T}\leftrightarrow {\cal T}_c$ to punctures with $U(1)$ flavor symmetry. Let us first look at the left hand side $\cal T$. After Higgsing the punctures, as outlined above, the theories $T_N$ and ${\tilde T}_N$ are transformed into the bifundamentals $(q,\tilde q)$ and $(\mathfrak{q},\tilde{\mathfrak{q}})$ respectively. The quarks $q$ (antiquarks $\tilde q$) are in the bifundamental (anti-bifundamental) of $SU(N)_B\times SU(N)_g$ and the quarks $\mathfrak{q}$ (antiquarks $\tilde{\mathfrak{q}}$) are in the bifundamental (anti-bifundamental) of $SU(N)_C\times SU(N)_g$. 
In addition, the Higgsing also produces $2N_{\mn}$ free chiral multiplets. All in all, we get the  SQCD $\cG$ with the superpotential \begin{equation}
W= c\tr (q\tilde q)_g (\mathfrak{q}\tilde{\mathfrak{q}})_g =
c\left[(q_{i\alpha}{\tilde{\mathfrak{q}}}^{k\alpha})({\tilde q}^{i\beta}\mathfrak{q}_{k\beta})-\frac{1}{N} (q_{i\gamma}{\tilde q}^{i\gamma})(\mathfrak{q}_{k\gamma}{\tilde{\mathfrak{q}}}^{k\gamma})\right],\label{superpotentialUx}
\end{equation}
where $\alpha, \beta$ are the gauge indices, and $i$ and $k$ are the $\SU(N)_{B}$ and $\SU(N)_{C}$ indices respectively.

On the dual side ${\cal T}_c$, the Higgsing produces the same number of free chiral multiplets and the $\cN{=}1$ $SU(N)$ SQCD with $N_f=2N$ with gauge singlet fields $M_B$ and $M_C$ in the adjoint representations of $SU(N)_B$ and $SU(N)_C$. We label by $p(\tilde p)$ and $\mathfrak{p} (\tilde{\mathfrak{p}})$ the quarks (anti-quarks) transforming in the bifundamental (anti-bifundamental) representation of $SU(N)_B\times SU(N)_{\hat g}$ and $SU(N)_C\times SU(N)_{\hat g}$ respectively. This is the theory $\cG_c$. The superpotential is
\be\label{superpotentialUc}
 W=\frac1c \tr(p \tilde p)_g (\mathfrak{p}\tilde{\mathfrak{p}})_g+ \tr  (p \tilde p)_B M_B+ \tr (\mathfrak{p}\tilde{\mathfrak{p}})_C M_C.
\ee
The subscript $X$ of the bilinear indicates that it transforms in the adjoint representation of $SU(N)_X$. 
This theory $\cU_c$ is the standard Seiberg dual of $\cU$. Indeed, 
 the first term of \eqref{superpotentialUx}  becomes a mass term for the mesons transforming in the bifundamental representation of $SU(N)_B\times SU(N)_C$ and the second term becomes the mass term for the singlet mesons. Integrating them out, we are only left with the mesons transforming in the adjoint representations of $SU(N)_B$ and $SU(N)_C$, together with the superpotential  \eqref{superpotentialUc}.
 
%%%%%%%%%%%%%%%
\subsubsection{New duality: $\cG\leftrightarrow \cG_s$}\label{newdual}
Let us now perform Higgsing at the punctures $A$ and $D$ on both sides of the duality ${\cal T}\leftrightarrow {\cal T}_s$. 
Higgsing the theory $\cal T$ produces the $\cN{=}1$ SQCD with $2N_\mn$ free chiral multiplets. 
Let us analyze what happens to the puncture $A$ on the $\cT_s$ side; the puncture $D$ behaves completely similarly.
We also present the analysis so that it applies to a general nilpotent vev.

The vev $\vev{\mu_A}=\rho_\mn(\sigma^+)$ is mapped to the vev $\vev{M_A}=\rho_\mn(\sigma^+)$. 
The superpotential of ${\cal T}_s$ becomes
\bea\label{superpotential-newdual} 
W =     \tr \rho_\mn(\sigma^{+}) {\hat \mu}_{A} + \tr M_{A} {\hat \mu}_{A}.
\eea
$M_{A}$ is now the fluctuation from the new vacuum. 
This type of deformation was studied in \cite{Heckman:2010qv}, we will follow the analysis there. The first term of the superpotential picks out a particular component of $\hat{\mu}_{A}$. Recall that the adjoint representation is decomposed into $\SU(2)$ representations as in \eqref{dec}, and the components of ${\hat \mu}_A$ can be written as ${\hat \mu}_{j,m,k}$ where $m=-j,-j+1,\ldots, j-1,j$ and $k=1,\ldots, \dim R_j$. 
The same  decomposition is applied to the components of $M_A$. 

The superpotential \eqref{superpotential-newdual} is
\be
W={\hat \mu}_{1,-1,1}+\sum_{j,m,k} M_{j,-m,k} {\hat \mu}_{j,m,k}.
\ee
Let $\CF_{0}$ be (the generator of) the original $\U(1)_{\CF}$. The first term apparently breaks this $U(1)_{\CF}$ symmetry because ${\CF_{0}}({\hat \mu}_{A})=+2$. 
Instead, the preserved combination is 
\be\label{Fshift}
\CF=\CF_0+2\rho_\mn(\sigma^3).
\ee 
Also, at the fixed point, the superpotential \eqref{superpotential-newdual} needs to be marginal. As $R_0({\hat \mu})=1$, we also need to redefine $U(1)_R$ symmetry so that $R({\hat \mu}_{1,-1,1})=2$. The new superconformal R symmetry is therefore
\be\label{Rshift}
R=R_0-\rho_\mn(\sigma^3).
\ee
The $SU(N)_A$ flavor symmetry is broken down to a subgroup. This results from the non-conservation
\be
{\bar D}^2(J_A)_{j,m,k}=\delta W={\hat \mu}_{j,m-1,k}.
\label{konishi}
\ee
This equation also means that the $\cN{=}1$ superconformal multiplet of current component $(J_A)_{j,m,k}$ recombines with the multiplet of ${\hat \mu}_{j,m-1,k}$ to become non-BPS. The remaining protected multiplets are those of $(J_A)_{j,-j,k}$ and ${\hat \mu}_{j,j,k}$. 
The currents $(J_A)_{0,0,k}$ are the generators of the flavor symmetry and others correspond to semi-conserved current multiplets. The $M_A$ fields, previously coupled to ${\hat \mu}_{j,m,k}$ for $m\neq j$, decouple\footnote{The authors thank Yu Nakayama for a useful discussion on this point.}. The number of such free fields is $N_\mn$.  
The same analysis goes through at puncture $D$ and results in $N_\mn$ more free fields. The number of free fields coming from the punctures $A$ and $D$ both matches the total number of free fields on the original side of the duality. The theory that we end up with on the dual side has
\bea
\CF
=     \CF_{0} + 2 \rho^{A}_\mn(\sigma^{3}) - 2 \rho^{D}_\mn(\sigma^{3}), ~~~
R
=     R_{0} - \rho^{A}_\mn(\sigma^{3}) - \rho^{D}_\mn(\sigma^{3}),
\label{Usshift}
\eea
and the gauge singlet fields $(M_A)_{j,-j,k}$ and $(M_D)_{j,-j,k}$ coupled to it through the superpotential
\be
W=\sum_{j,k} (M_A)_{j,-j,k}{(\hat \mu_A)}_{j,j,k}+\sum_{j,k}  (M_D)_{j,-j,k}{(\hat \mu_D)}_{j,j,k}.
\ee
This is our new dual to $\cN{=}1$ SQCD. 
The adjoint representation of $\SU(N)_{A}$ is decomposed into $N+1$ irreducible representations, giving $N+1$ fields $(M_A)_{j,-j,k}$.  Therefore, in total there are $2(N^{2} + N)$ gauge single fields.

%%%%%%%%%%%%%%%%%%%%%%%%%%%%%%%%%%%%%%%  
\subsubsection{Operator matching}
The duality $\cG \leftrightarrow \cG_s$ involves the following matching of gauge invariant chiral operators. 
Using the notations in section \ref{sd}, the mesons and the baryons in the SQCD $\mathcal{U}$ are
\bea \label{chiralringU}
m_{i}^{j}
=     q_{i} \tilde{q}^{j},~~~
\tilde{m}_{k}^{\ell}
=     \mathfrak{q}_{k} \tilde{\mathfrak{q}}^{\ell},~~~
\mathfrak{m}_{i}^{k}
=     q_{i} \tilde{\mathfrak{q}}^{k},~~~
\tilde{\mathfrak{m}}_{k}^{i}
=     \mathfrak{q}_{k} \tilde{q}^{i},~~~~~
b^{(k)}
=     \epsilon (q^{N-k} \mathfrak{q}^{k}).
\eea
In the last equation the gauge indices are contracted by $\epsilon$ tensor of $SU(N)$ gauge symmetry. One can also construct  the anti-baryons $\epsilon({\tilde q}^{N-k} {\tilde{\mathfrak q}}^k)$, whose matching will work in the same way as baryons. The $\U(1)$ charges of the chiral operators are summarized in Table 1.
\begin{table}[t]
\begin{center}
\begin{tabular}{|c|c|c|c|c|}
\hline 
& $U(1)_{{\cal F}}$ & $U(1)_{R}$ & $U(1)_{A}$ & $U(1)_{D}$ \tabularnewline
\hline 
\hline 
$m$ & $-2$ & $1$ & $0$ & $0$ \tabularnewline
\hline 
$\tilde{m}$ & $+2$& $1$ & $0$ & $0$ \tabularnewline
\hline
$\mathfrak{m}$ & $0$ & $1$ & $+1$  & $-1$ \tabularnewline
\hline
$\tilde{\mathfrak{m}}$ & $0$ & $1$ & $-1$ & $+1$  \tabularnewline
\hline 
$b^{(k)}$ & $2k - N$ & $\frac{N}{2}$ & $N-k$ & $k$ \tabularnewline
\hline 
\end{tabular}
\label{table:mesonbaryon}
\caption{The $\U(1)$ charges of the gauge-singlet fields in $\mathcal{U}$.}
\end{center}
\end{table}
  
The $\mu$ operators coming from $T_N$ and ${\tilde T}_N$ of $\cG_s$ do not exist in the chiral ring due to superpotential coupling with singlet fields $M$. The mesons $m$ and $\tilde m$ in the list \eqref{chiralringU} map to the singlet fields $M_B$ and $M_C$ of $\cG_s$. The trifundametal operator $Q_{ijk}=:Q_{(1)}$ of the component $T_N$ theory on the dual side splits into $N$ operators transforming in the bifundamental representation of $SU(N)_B \times SU(N)_g$ after Higgsing the $SU(N)$ symmetry at  the $A$ puncture down to $U(1)$. These $N$ components form one $N-1$ dimensional representation $Q_{(1)\frac{N-2}{2},m}$ $(-\frac{N-2}{2}\leq m \leq \frac{N-2}{2})$ and one singlet representation $Q_{(1)0,0}$ of the embedded $SU(2)$. Each component of $Q_{(1)}$ carries the $U(1)_R$ and $U(1)_{\CF}$ charge that is determined by the eq. \eqref{Usshift}. Similar analysis is applied to the anti-trifundamental $Q^{ijk}=:Q_{(N-1)}$. From these shifted charges, we conclude the mapping
\be
{\mathfrak m} \leftrightarrow Q_{(1)\frac{N-2}{2},\frac{N-2}{2}}{\tilde Q}_{(N-1)\frac{N-2}{2},-\frac{N-2}{2}} \qquad 
{\tilde {\mathfrak m}} \leftrightarrow Q_{(N-1)\frac{N-2}{2},-\frac{N-2}{2}}{\tilde Q}_{(1)\frac{N-2}{2},\frac{N-2}{2}}.
\ee
The baryons $b^{(0)}$ and $b^{(N)}$ are mapped to $(M_A)_{\frac{N-2}{2},-\frac{N-2}{2}}$ and $(M_D)_{\frac{N-2}{2},-\frac{N-2}{2}}$ respectively. Here we have split the adjoint representation of $SU(N)$ into irreducible representations of embedded $SU(2)$
and picked up the appropriate component. The other baryons $b^{(k)}$ are expected to match the appropriate representation appearing in the product $Q_{(N-k)}{\tilde Q}_{(k)}$. In this section, we have described the matching of a few prominent chiral operators on both sides. In section \ref{sec:index} we will match the superconformal index of $\cG$ and $\cG_s$.

%%%%%%%%%%%%%%%% section 3 %%%%%%%%%%%%%%%%%%%%%%%%%%%%%%%%%%%%%%%%%%%
\section{Checks of the anomaly}
\label{sec:anomaly}
In this section, we compute the 't Hooft anomaly coefficients on both sides of the duality discussed in the previous section. Their agreement gives a non-trivial check for our proposal. We first consider the duality $\CT \leftrightarrow \CT_s$ and then study the duality $\cU\leftrightarrow \cU_s$.

\subsection{Known facts}\label{known}
Let us first summarize known facts. 
The central charges $a$ and $c$ of an $\CN{=}1$ SCFT are linear combinations of the 't Hooft anomalies
$\tr R$ and $\tr R^{3}$.
For an $\cN{=}2$ theory the anomalies satisfy the relations
\begin{equation}
\tr R_{\CN{=}2}
= \tr R_{\CN{=}2}^{3}
=  2(n_{v} - n_{h}), \qquad
\tr R_{\CN{=}2} I_{3}^{2}
=  \frac{1}{2} n_{v}
\label{acnvnh}
\end{equation}
where $n_v$ and $n_h$ are effective number of  vector multiplets and hypermultiplets, respectively.

The numbers $n_{v,h}$ of the $T_N$ theory and its cousins obtained by closing the punctures are in turn given by \begin{equation}
n_{v}
=  - \left( \frac{4}{3} N (N^{2}-1)
+ N-1 \right)+\sum_{X} n_{v}(\rho_{\Lambda_{X}}),\qquad
n_{h}
=  - \frac{4}{3} N (N^{2}-1)+\sum_{X} n_{h} (\rho_{\Lambda_{X}}),
\end{equation}
where $n_{v}(\rho_{\Lambda})$ and $n_{h}(\rho_{\Lambda})$ are the contribution of the puncture specified by $\Lambda$. They are given by \cite{Chacaltana:2012zy}
\bea
n_{v}(\rho_{\Lambda})
&=& \frac{2}{3} N (N^{2} -1 ) - 4 \rho_{\mathfrak{g}} \cdot
\rho_{\Lambda}(\sigma_{3})
+ \frac{1}{2} (N-1 - \dim \mathfrak{g}_{0}),
\nonumber \\
n_{h}(\rho_{\Lambda})
&=& \frac{2}{3} N (N^{2} -1 ) - 4 \rho_{\mathfrak{g}} \cdot
\rho_{\Lambda}(\sigma_{3})
+ \frac{1}{2} (\dim \mathfrak{g}_{1/2}),
\eea
where $\dim \mathfrak{g}_{0}$ and $\dim \mathfrak{g}_{1/2}$ are the numbers of the representations with even spin and with odd spin, respectively, in the decomposition.
For the maximal puncture $\rho_\text{max}$ we have \bea
n_{v}(\rho_{\text{max}})
= \frac{1}{6} N (N-1)(4N+1), ~~~
n_{h}(\rho_{\text{max}})
= \frac{2}{3} N(N^{2}-1).
\label{nvmax}
\eea 
For the puncture $\mn=(N-1,1)$ we have \begin{equation}
n_{v}(\rho_\mn) = N^{2} - 1,\qquad n_{h}(\rho_\mn)= N^{2}
\end{equation} from which we deduce $n_{v} = 0$ and $n_{h} = N^{2}$ for a three-punctured sphere with two maximal punctures and one puncture $\mn$. This reproduces the anomalies of a bifundamental.

The flavor central charge is defined by \cite{Anselmi:1997am,Anselmi:1997ys}
\bea
K_{\SU(N)} \delta^{ab}
= -3 \tr R T^{a} T^{b},
\label{defflavor}
\eea
where $T^{a}$ are generators of the $\SU(N)$ flavor symmetry. 
We normalize the flavor central charge such that $N$ free chiral multiplets give $K_{\U(N)} = \frac{1}{2}$. 
We also normalize the quadratic Casimirs as $C_{2}(\square) = \frac{1}{2}$ and $C_{2}({\rm adj}) = N$ for $\SU(N)$.

In an $\CN{=}2$ theory, the flavor symmetry central charge is equivalently given by 
\bea
k \delta^{ab}
= - 2 \tr R_{\CN{=}2} T^{a} T^{b},
\label{TNflavor}
\eea
For the $T_N$ theory and its cousins obtained by closing the punctures, the central charge of the flavor symmetry of the puncture $\Lambda$ is given by \cite{Chacaltana:2012zy} 
\bea
k \delta^{ab}
= 2 \sum_{j} \tr_{R_{j}} T^{a} T^{b},
\label{TNflavorU(1)}
\eea
where $R_{j}$ are the representations of the flavor symmetry appeared in the decomposition \eqref{dec}. 
Let $\Lambda^{T} =(s_{1}, s_{2}, \ldots, s_{n_{1}})$ be the dual partition to $\Lambda$. We also define $r_{i} = s_{i} - s_{i+1}$. 
The flavor symmetry of the puncture $\Lambda$ is $S[\prod_{i=1}^{n_{1}} \U(r_{i})]$. The flavor central charge is then given by
\bea
k_{\SU(r_{i})}
= 2 \sum_{j \leq i} s_{j}, \label{orgflavor}
\eea
for the $\SU(r_{i})$ subgroup. In the case of the maximal puncture where $s_{1} = N$, we get the flavor central charge $k_{\SU(N)} = 2N$. The flavor central charge of $\U(1)$ subgroup can be computed in turn by using \eqref{TNflavorU(1)}.

%%%%%%%%%%%%%%%%%%%%%%%%%%%%%%%
\subsection{Dualities of coupled $T_N$ theories}

When one starts from an $\cN{=}2$ theory and deform it to an $\cN{=}1$ theory,  we can combine \eqref{chargerelation1} and \eqref{acnvnh} to obtain the anomaly of the IR R-symmetry in terms of $n_v$ and $n_h$ of the $\cN{=}2$ theory:
\bea
\tr R
= n_{v} - n_{h}, ~~~
\tr R^{3}
= n_{v} - \frac{1}{4} n_{h}.
\label{ele}
\eea
Since the gauge singlet fields $M$ have R-charge one, they do not contribute to $\tr R$ and $\tr R^{3}$. Therefore, the anomalies $\tr R$ and $\tr R^3$ are clearly the same for $\cT$ and $\cT_s$. Let us discuss other anomalies. 

We focus on a puncture with $\SU(N)$ flavor symmetry and compute its flavor central charge.
Again the gauge singlets do not contribute on the dual side, thus we only need to consider the contribution of the $T_{N}$ theory. Due to \eqref{chargerelation1}, the central charge is written as $K_{\SU(N)} \delta^{ab} = - \frac{3}{2} \tr R_{\CN{=}2} T^{a} T^{b}$. Thus we get $K_{\SU(N)} = \frac{3}{2} N$, by using \eqref{TNflavor}, on the both sides of the duality.

Next let us compute the anomaly coefficient $\tr \CF T^{a} T^{b}$. The contribution to this anomaly from the $T_{N}$ theory with sign $\sigma$ can be computed by using \eqref{chargerelation2} and \eqref{TNflavor}
\bea
\tr \CF T^{a} T^{b}
= - \sigma N \delta^{ab}.
\eea
On the dual side, by adding the contribution of $M$ which has $\U(1)_{\CF}$ charge $-2\sigma$, we obtain $(\sigma N + (-2 \sigma) C_{2}({\rm adj}))\delta^{ab}$, which agrees with that of the original theory. 

The anomaly coefficient $\tr T^{a} T^{b} T^{c}$ can be easily seen to be the same on both sides of the duality as the $M$ field on the dual side transforms in the adjoint representation and hence doesn't contribute to $\tr T^{a} T^{b} T^{c}$. The anomaly coefficients $\tr \CF^{2} R$  also match because the $M$ fields do not contribute. The anomaly coefficients $\tr \CF^{3}$, $\tr \CF$ and $\tr \CF R^{2}$ agree because $M_{A}$, $M_{B}$ and $M_{C}$, $M_{D}$ have opposite $\U(1)_{\CF}$ charge and hence their contribution is cancelled.

%%%%%%%%%%%%%%%%%%%%%%%%%%%%%%%%%%%%%%%%%%%
\subsection{Dualities of SQCD}
Let us turn to the duality $\mathcal{U} \leftrightarrow \mathcal{U}_{s}$. This duality is obtained by Higgsing both sides of the duality ${\cal T}\leftrightarrow {\cal T}_s$ with a vev $\rho_{\mn}$. The analysis in this section is general and can be applied to Higgsing the symmetry with a vev $\rho_\Lambda$ for any $\Lambda$.

\paragraph{Central charges}
Since the contribution from the $\CN{=}1$ vector multiplet is the same on the both sides, we focus on the contribution of the component $T_{N}$ theory in $\cal T$ with $\sigma=+1$. We Higgs the flavor symmetry at puncture $A$ with the vev $\rho_{\Lambda}$. The anomalies $\tr R$ and $\tr R^3$ of this component is obtained by plugging 
 $n_{v}$ and $n_{h}$  listed in section \ref{known} into the formula \eqref{ele}.

On the dual side, as discussed in section \ref{newdual}, the IR R symmetry is shifted as \eqref{Usshift}. The $\tr R$ is not affected by the shift, but $\tr R^{3}$ becomes
\bea
\tr R^{3}
= \tr R_{0}^{3} + 3 \tr R \rho_\Lambda(\sigma^{3})^{2},
\eea
where the first term is the anomaly coefficient of the unHiggsed $T_{N}$ theory. The second term gets related to $\tr R_{\CN{=}2} T^{a} T^{b}$, it is given as  $\frac{3}{2} I_{{\Lambda}} \tr R_{\CN{=}2} T^{a} T^{b} = - \frac{3}{2} N I_{{\Lambda}} \delta^{ab}$, where $I_{{\Lambda}}$ is the embedding index associated to $\rho_\Lambda:SU(2)\to SU(N)$.
\bea
I_{{\Lambda}}
&=& \frac{1}{6} \sum_{\alpha=1}^{\ell} n_{\alpha} (n_{\alpha}^{2}-1).
\eea
Furthermore, due to \eqref{konishi}, the singlets $M_{j,m,k}$ with $m\neq j$ become free and decouple. The remaining gauge singlet fields are $M_{j,-j,k}$ with R-charge $1+j$, $\CF$ charge $-2-2j$. Their contribution to $\tr R$ is
\bea
c_{1}
= \sum_{\alpha=1}^{\ell} \sum_{j=1}^{n_{\alpha}-1} j
+ 2 \sum_{\alpha< \beta} \sum_{k=1}^{n_{\beta}}
\left(\frac{n_{\alpha} + n_{\beta} - 2k}{2} \right).
\eea
and that to $\tr R^{3}$ is
\bea
c_{3}
= \sum_{\alpha=1}^{\ell} \sum_{j=1}^{n_{\alpha}-1} j^{3}
+ 2 \sum_{\alpha< \beta} \sum_{k=1}^{n_{\beta}}
\left(\frac{n_{\alpha} + n_{\beta} - 2k}{2} \right)^{3}.
\eea
Thus, the 't Hooft anomaly coefficients are
\bea
\tr R^{3}
= \frac{N^{3}}{2} - \frac{3N^{2}}{2} + 1 - \frac{3}{2} N I_{{\Lambda}} + c_{3},
\qquad
\tr R
= - \frac{3N^{2}}{2} + \frac{N}{2} +1 + c_{1}.
\eea

By explicit evaluations we find that the expressions for $\cU$ and $\cU_s$ always give the same results. 

%%%%%%%%%%%%%%%%
\paragraph{Flavor anomalies}
Let us next consider the flavor central charge, focusing on the puncture Higgsed by the vev $\rho_{\Lambda}$, whose flavor symmetry is $S [\prod_{i=1}^{n_{1}} \U(r_{i})]$. On the original side, the flavor central charge of $\SU(r_{i})$ subgroup is 
\bea
K_{\SU(r_{i})}
= \frac{3}{2} \sum_{j \leq i} s_{j},
\label{flavor}
\eea 
which immediately follows from \eqref{orgflavor}.

The dual theory has the same flavor symmetry as the original one. The flavor central charge of the dual theory can be written as
\bea
K_{\SU(r_{i})}^{{\rm dual}} \delta^{ab}
= - 3 \tr R t^{a} t^{b}
= I_{\SU(r_{i}) \subset \SU(N)} K_{\SU(N)} + K_{M},
\label{flavordualSU(r)}
\eea
where $t_a$ are the generators of $SU(r_i)$, $K_{\SU(N)} = \frac{3N}{2}$ is the flavor central charge of the $T_{N}$ theory, $I_{\SU(r_{i}) \subset \SU(N)}$ is the embedding index of $\SU(r_{i})$ in $\SU(N)$, and $K_{M}$ is the contribution of the surviving $M$ fields. Let us first compute the embedding index. The partition $\Lambda = (n_{1}, n_{2}, \ldots, n_{\ell})$ is labeled by the $\alpha$ index ($\alpha = 1, \ldots, \ell$). Then, let us consider the following case: $\alpha = q_{i}$ is the leftmost entry with the value $n_{q_{i}}$: $n_{1} \geq n_{2} \geq \ldots \geq n_{q_{i}-1} > n_{q_{i}} = n_{q_{i}+1} = \ldots = n_{q_{i} + r_{i} -1} > n_{q_{i} + r_{i}} \geq \ldots$. In this notation, the embedding index is  simply $I_{\SU(r_{i}) \subset \SU(N)} = n_{q_{i}}$.

To calculate $K_{M}$, we need to find the gauge singlets $M$ which are in nontrivial representations of the flavor symmetry $\SU(r_{i})$. By the embedding $\Lambda$, we decompose $M$ transforming originally in the adjoint representation of $\SU(N)$ into the following parts:
\bea
M
= \left( \begin{array}{ccc}
A & B & C \\
B' & D & E \\
C' & E' & F
\end{array}
\right),
\eea
where $A$ is an $\hat{n} \times \hat{n}$ matrix ($\hat{n} = \sum_{\alpha=1}^{q_{i} -1} n_{\alpha}$), $D$ is an $(n_{q_{i}} r_{i}) \times (n_{q_{i}} r_{i})$ matrix, and $F$ is an $\tilde{n} \times \tilde{n}$ matrix ($\tilde{n} = \sum_{\alpha = q_{i} + r_{i}}^{\ell} n_{\alpha}$). The generators of the flavor symmetry $\SU(r_{i})$ is expressed as
\bea
\left(
\begin{array}{ccc}
0~~~ & 0 & ~~~0 \\
0~~~ & \mathfrak{su}(r_{i}) \otimes n_{q_{i}} & ~~~0 \\
0~~~ & 0 & ~~~0
\end{array}
\right).
\eea
Thus, the only parts which are charged under this symmetry are $D$, $B$, $B'$, $E$ and $E'$. Since $B'$ and $E'$ are similar to $B$ and $E$, we only consider $D$, $B$ and $E$.

First of all, $D$ is further decomposed into $r_{i}^{2}$ small blocks, each of which is an $n_{q_{i}} \times n_{q_{i}}$ matrix. We can think of this block as in ${\rm adj} \oplus 1$ representation of $\SU(r_{i})$. Then, we need to know which components of the $n_{q_{i}} \times n_{q_{i}}$ matrix survive the Higgsing. As in the first term in \eqref{dec}, the $n_{q_{i}} \times n_{q_{i}}$ matrix is decomposed into the $\SU(2)$ representation as $n_{q_{i}}^{2} = \oplus_{k=1}^{n_{q_{i}}-1} V_{k} \oplus V_{0}$. The bottom component of each $V_{k}$ representation has $R$ charge $1+k$ and corresponds to the surviving components of $M$. Therefore, the contribution to $K_{M}$ from $D$ component is
\bea
K_{M(D)}
= - 3 \sum_{k=1}^{n_{q_{i}}-1} k \,C_2 ({\rm adj})_{\SU(r_{i})}
= - \frac{3}{2} r_{i} (n_{q_{i}} - 1) n_{q_{i}}.
\eea

Next, let us consider the $B$ part. This is an $(r_{i} n_{q_{i}}) \times \hat{n}$ matrix. This can be decomposed into $r_{i}$ blocks, each of which is an $n_{q_{i}} \times \hat{n}$ matrix. They form the fundamental representation of $\SU(r_{i})$. Then, we have to consider the further decomposition of $n_{q_{i}} \times \hat{n}$ into $\SU(2)$ representations: $n_{q_{i}} \times \hat{n} \rightarrow \oplus_{\alpha = 1}^{q_{i}-1} \oplus_{k=1}^{n_{q_{i}}} V_{\frac{n_{\alpha} + n_{q_{i} - 2k}}{2}}$. Therefore, the contribution of $B$ (and $B'$) to the flavor central charge is
\bea
K_{M(B, B')}
= - 6 \sum_{\alpha = 1}^{q_{i}-1} \sum_{k=1}^{n_{q_{i}}}
\frac{n_{\alpha} + n_{q_{i}} - 2k}{2}
C_{2}({\bf r_{i}})
= - \frac{3 n_{q_{i}}}{2} \left( \sum_{\alpha=1}^{q_{i}-1} n_{\alpha} -
 q_{i} + 1 \right).
\eea
Similarly, the contribution of $E$ and $E'$ can be computed as
\bea
K_{M(E, E')}
&=& - 6 \sum_{\alpha = q_{i} + r_{i}}^{\ell} \sum_{k=1}^{n_{\alpha}}
\frac{n_{q_{i}} + n_{\alpha} - 2k}{2} C_{2}({\bf r_{i}})
= - \frac{3}{2} (n_{q_{i}} - 1 ) \sum_{\alpha = q_{i} + r_{i}}^{\ell}
n_{\alpha}.
\eea
By adding all the contributions we obtain the flavor central charge of the dual theory $K_{\SU(r_{i})}^{{\rm dual}} = \frac{3}{2} \sum_{j \leq i} s_{i}$, which is equivalent to that of the original theory \eqref{flavor}.

So far, we have not considered the central charge of a $\U(1)$ flavor symmetry. To illustrate the agreement, we consider the case with $\Lambda = \mn := (N-1, 1)$ whose flavor symmetry is just $\U(1)$. By the decomposition of the adjoint representation, the upper-right off-diagonal part of the adjoint of $\SU(N)$ has charge $N$, and the lower-left off-diagonal part has charge $-N$, under this $\U(1)$. These are in spin $\frac{N-2}{2}$ representations of $\SU(2)$. Note that the generator of this $\U(1)$ in $\SU(N)$ is $\diag (1, 1, \ldots, 1, -(N-1))$.

On the original side, by using \eqref{TNflavorU(1)} we obtain
\bea
K_{\U(1)} \delta^{ab}
= \frac{3}{4} k_{\U(1)} \delta^{ab}
= 3 N^{2} \delta^{ab}.
\eea
Indeed, we can get the same answer by directly considering $-3 \tr R T^{a} T^{b}$ of the quarks $q$ and the anti-quarks $\tilde{q}$ whose R and $\U(1)$ charges are $\frac{1}{2}$ and $\pm 1$ respectively.

On the dual side, we again consider
\bea
K^{{\rm dual}}_{\U(1)}
= I_{\U(1) \subset \SU(N)} K_{\SU(N)} + K_{M},
\label{flavordualU(1)}
\eea
where the embedding index is $I_{\U(1) \subset \SU(N)} = 2 N (N-1)$. Since the surviving gauge singlet fields are the bottom components of two spin $\frac{N-2}{2}$ representations, with R-charge $1 + \frac{N-2}{2}$, thus $K_{M} = - 3 \cdot 2 (\frac{N-2}{2}) N^{2} = 3 (N^{3} - 2N^{2})$. Thus, we obtain the same result as the original $\U(1)$ central charge $K_{\U(1)}^{{\rm dual}}= 3N^{2}$.

The anomaly coefficient $\tr \CF T^{a} T^{b}$ can be computed in a similar fashion. On the original side, by using \eqref{chargerelation2}, this is related with the flavor central charge \eqref{flavor}. On the dual side, the similar computation as \eqref{flavordualSU(r)} and \eqref{flavordualU(1)} including the contribution of the gauge singlet fields leads to the nontrivial agreement. We do not consider the other anomaly coefficients here.

%%%%%%%%%%%%%%%% section 4 %%%%%%%%%%%%%%%%%%%%%%%%%%%%%%%%%%%%%%%%%%%
\section{Checks of the index}
\label{sec:index}

Let us start by defining the superconformal index for both $\cN{=}1$ and $\cN{=}2$ SCFTs. The $\cN{=}1$ index is defined as
\be
{\cal I}^{{\cal N}=1}(z;p,q,\xi)=\mbox{Tr}(-1)^{F}p^{j_{1}-j_{2}+\frac{R}{2}}q^{j_{1}+j_{2}+\frac{R}{2}}\xi^{-\frac{{\cal F}}{2}}z^{Q}
\ee
Here, $j_{1}$ and $j_{2}$ are spins with respect to $SO(4)\simeq SU(2)_{1}\times SU(2)_{2}$
rotational symmetry, $R$ is the $U(1)$ R-charge. If the $\cN{=}1$ theory has a flavor symmetry we can also incorporate it by turning on the fugacity $z$ that couples to flavor symmetry Cartan generators. In addition to non-abelian flavor symmetries, the class of theories considered in this paper also possesses the $U(1)_{\CF}$ flavor symmetry. We use a special variable $\xi$ for its fugacity. The $\cN{=}2$ index is defined as
\be
{\cal I}^{{\cal N}=2}(z;p,q,t)=\mbox{Tr}(-1)^{F}p^{j_{1}-j_{2}+\frac{R_{{\cal N}=2}}{2}}q^{j_{1}+j_{2}+\frac{R_{{\cal N}=2}}{2}}t^{I_{3}-\frac{R_{{\cal N}=2}}{2}}z^{Q}.
\ee
In this expression $R_{{\cal N}=2}$ and $I_{3}$ are $U(1)_{R}$ and $SU(2)_{R}$ R symmetries of the ${\cal N}=2$ theory. 

\subsection{Known facts}

Consider the $\cN{=}2$ theory constructed by coupling two $T_N$ theories with $\cN{=}2$ vector multiplet. After integrating out the adjoint chiral field in the $\cN{=}2$ vector multiplet, we obtain the $\cN{=}1$ theory $\cal T$ that is central to this paper. The $U(1)_R$ and $U(1)_{\CF}$ charge is determined by the following combination of $U(1)_R$ and $SU(2)_R$ charges of the $T_N^{(i)}$ theories $(i=1,2)$ involved.
\be\label{RFchargemap}
R=\sum_i \frac{1}{2}R_{\cN{=}2}^{(i)}+I_3^{(i)}\qquad \CF=\sum_i \sigma_i(R_{\cN{=}2}^{(i)}-2I_3^{(i)}),
\ee
where $\sigma_1=+1$ and $\sigma_2=-1$. From now on, we define $\sigma$ to be the \emph{sign} of $T_N$ and label the individual $T_N$ blocks with their sign as $T_N^\sigma$. If we consider $\cN{=}1$ charges of the parent $\cN{=}2$ theory (without integrating out the adjoint chiral field), $\sigma_1=\sigma_2=+1$. These conclusions may be rephrased as: $T_N$ theories with opposite signs are coupled using the $\cN{=}1$ vector multiplet and $T_N$ theories of the same sign are coupled using $\cN{=}2$ vector multiplet.

From eq. \eqref{RFchargemap}, it is easy to write the $\cN{=}1$ index of $T_N^\sigma$ in terms of $\cN{=}2$ index of the $T_N$ theory
\be
I_{T_N^\sigma}(\,\cdot\,;p,q,\xi)=I_{T_N}^{\cN{=}2}(\,\cdot\,;p,q,\xi^\sigma \sqrt{pq}).
\ee
Here, $\cdot$ stands for all the fugacities corresponding to $SU(N)^3$ flavor symmetry.  The ${\cal N}=2$ index of the $T_{N}$ theories has been computed in \cite{Gaiotto:2012xa}. It is most conveniently expressed in terms of symmetric function of $N$ variables $\psi_{\lambda}(a_{i};p,q,t)$. When not ambiguous, we will frequently drop the arguments $p,q,t$ :
\be\label{N=2TNindex}
{\cal I}_{T_{N}}^{{\cal N}=2}(\bfa,\bfb,\bfc)=\sum_{\lambda}C_{\lambda}\psi_{\lambda}(\bfa)\psi_{\lambda}(\bfb)\psi_{\lambda}(\bfc).
\ee
We have denoted the set of $N$ variables $\{a_i\}$ by $\bfa$ and so on. The functions $\psi_{\lambda}(\bfa)$ are orthonormal under the integration with respect to ${\cal N}=2$ vector multiplet measure $[d\bfa]{\cal I}_{V}^{{\cal N}=2}(\bfa)$ where $[d\bfa]$ is the usual Haar measure for $SU(N)$. They satisfy an identity
\be\label{niceid}
\psi_{\lambda}(\bfa;p,q,t)=\mbox{PE}\left(\frac{t-pq/t}{(1-p)(1-q)}\chi_{\mbox{adj}}(\bfa)\right)\psi_{\lambda}(\bfa;p,q,\frac{pq}{t}).
\ee
For our purposes, it is most convenient to define the functions $\Psi_{\lambda}(\bfa)$,
\be
\mbox{\ensuremath{\psi}}_{\lambda}(\bfa;p,q,t)=:K(t,\bfa)\Psi_{\lambda}(\bfa;p,q,t),\qquad K(t,\bfa):=\mbox{PE}\left(\frac{t-pq}{(1-p)(1-q)}\chi_{\text{adj}}(\bfa)\right).
\ee
From eq. \eqref{niceid}, it follows that the new functions $\Psi_{\lambda}(\bfa;p,q,t)$ have a nice property that they are exactly symmetric under the exchange $t\leftrightarrow pq/t$ i.e. $\xi\leftrightarrow\xi^{-1}$. 

\paragraph{Higgsing}
When we Higgs a puncture to partition type $\Lambda$ in ${\cal N}=2$ theory, the adjoint representation of $SU(N)$ decomposes as $\oplus R_{j}\otimes V_{j}$ where $V_{j}$ is a spin $j$ representation of $SU(2)$ and $R_{j}$ is a representation of the commutant. We conjecture the following prescription to get the index of the theory with $\Lambda$ type puncture. We substitute
\be
\label{general-prescription}
\Psi_{\lambda}(\bfa)\to\Psi_{\lambda}(\bfu t^{\Lambda}),\qquad K(t,\bfa)\to K_{\Lambda}(t,\bfu):=\mbox{PE}\left(\sum_{j}\frac{t^{1+j}-pqt^{j}}{(1-p)(1-q)}\chi_{R_{j}}(\bfu)\right).
\ee
The meaning of substituting $\bfu t^\Lambda$ for the argument $\bfa$ of $\Psi_{\lambda}$ is perhaps most conveniently explained  through an example. Consider Higgsing of a maximal puncture of $SU(5)$ by a vev corresponding to the partition $\Lambda=(3,1,1)$. The flavor symmetry associated to this puncture is now $S[U(1)\times U(2)]$. In this case, the function $\Psi(a_1,\ldots, a_5; pq,t)$ is evaluated on,
\be
\bfa=\{a_1,a_2,a_3,a_4,a_5\}|_{\prod_{i=1}^{5}a_i=1} \longrightarrow \bfu t^\Lambda=\{t^2 u_1, u_1, t^{-2} u_1, u_2, u_3 \}|_{ u_1^3 u_2 u_3=1}.
\ee
This prescription generalizes the by-now standard formula in the Macdonald limit $p\to 0$ given in \cite{Gadde:2011uv} and slightly simplified in \cite{Mekareeya:2012tn}. The choice of the prefactor $K_{\Lambda}(t,a)$ in  \eqref{general-prescription} is motivated by the fact that it is the contribution from conserved and semi-conserved multiplets of spin $j$.  As we will see below, with this choice the superconformal index behaves nicely under our duality with arbitrary punctures. This can be seen as an overall check of consistency. We also checked the formula \eqref{general-prescription} in the limit $p=q=t$ where $\Psi_{\lambda}=\chi_\lambda$, against the index of the bifundamentals for $N=2,3,4$. We can now write the ${\cal N}=1$ index of $T_{N}^\sigma$ theory,
\be
{\cal I}_{T_{N}^\sigma}(\bfa,\bfb,\bfc)=\frac{K(\xi^{\sigma}\sqrt{pq},\bfa)K(\xi^{\sigma}\sqrt{pq},\bfb)K(\xi^{\sigma}\sqrt{pq},\bfc)}{K_{\varnothing}(\xi^{\sigma}\sqrt{pq})}\sum_{\lambda}\frac{\Psi_{\lambda}(\bfa)\Psi_{\lambda}(\bfb)\Psi_{\lambda}(\bfc)}{\Psi_{\lambda}((\xi^{\sigma}\sqrt{pq})^{\varnothing})}.
\ee
In addition to using the eq. \eqref{N=2TNindex}, we have also simplified the structure constant $C_{\lambda}$. The equation can be written much more compactly if we define $K_{\Lambda}(\xi^{\sigma}\sqrt{pq},\bfa)=:K_{\Lambda}^{\sigma}(\bfa)$ and $\Psi_{\lambda}((\xi^{\sigma}\sqrt{pq})^{\Lambda}\bfa)=:\Psi_{\lambda}^{\Lambda\sigma}(\bfa)$. 

\subsection{Dualities of coupled $T_N$ theories}
In this subsection we will show the equality of the superconformal index on both sides of the duality ${\cal T} \leftrightarrow {\cal T}_s$. The theory $\cal T$ obtained by coupling $T_{N}^{+}$ and $T_{N}^{-}$ theory with the ${\cal N}=1$ vector multiplet. Its index is
\be
{\cal I}_{{\cal T}}(\bfa,\bfb;\bfc,\bfd)=\oint[d{\bf z}]{\cal I}_{V}^{{\cal N}=1}({\bf z}){\cal I}_{T_{N}^{+}}(\bfa,\bfb,{\bf z}){\cal I}_{T_{N}^{-}}({\bf z},\bfc,\bfd).
\ee
The orthonormality of the functions $\psi_\lambda({\bf z})$ under the measure $[d{\bf z}]{\cal I}_{V}^{{\cal N}=2}({\bf z})$ results in the orthonormality of the functions $\Psi_\lambda({\bf z})$ under the measure $[d{\bf z}]{\cal I}_{V}^{{\cal N}=1}({\bf z})K(\xi\sqrt{pq},{\bf z})K(\xi^{-1}\sqrt{pq},{\bf z})$. Fortunately, this is exactly the measure that appears in the above integral. Using the orthonormality, we write
\be
{\cal I}_{{\cal T}}(\bfa,\bfb;\bfc,\bfd)=\frac{K^{+}(\bfa)K^{+}(\bfb)K^{-}(\bfc)K^{-}(\bfd)}{K_{\varnothing}^{+}K_{\varnothing}^{-}}\sum_{\lambda}\frac{\Psi_{\lambda}(\bfa)\Psi_{\lambda}(\bfb)\Psi_{\lambda}(\bfc)\Psi_{\lambda}(\bfd)}{\Psi_{\lambda}^{\varnothing+}\Psi_{\lambda}^{\varnothing-}}.
\ee
This expression was obtained as a TQFT correlator on four-punctured sphere in \cite{Beem:2012yn}.

On the other side of the duality, the theory ${\cal T}_s$ consists of a copy of $\cal T$ but with opposite $U(1)_{\cal F}$ charge. In addition it also has gauge singlet fields $M_X$ (for $X=A,B,C,D$) which transform in the adjoint representation of $SU(N)_X$ flavor symmetry. Among them, $M_A$ and $M_B$ have $\CF=-2$ and $M_C$ and $M_D$ have $\CF=+2$. The index of the first part is obtained from the index of $\cal T$ after swapping $(\bfa,\bfb)\leftrightarrow (\bfc,\bfd)$. The index of $M_X$ with $\CF=-2\sigma$ is
\be\label{Mindex}
M^{\sigma}(\bfa)=\mbox{PE}\left(\frac{\sqrt{pq}(\xi^\sigma-\xi^{-\sigma})}{(1-p)(1-q)}\chi_{\text{adj}}(\bfa)\right).
\ee
Putting everything together, the index of ${\cal T}_s$ is:
\be
{\cal I}_{{\cal T}_{s}}(\bfa,\bfb;\bfc,\bfd)=M^{+}(\bfa)M^{+}(\bfb)M^{-}(\bfc)M^{-}(\bfd){\cal I}_{{\cal T}}(\bfc,\bfd;\bfa,\bfb).
\ee
Indeed ${\cal I}_{{\cal T}}={\cal I}_{{\cal T}_{s}}$, thanks to the identity
\be\label{cid}
M^{\sigma}(\bfa)=K^{\sigma}(\bfa)/K^{-\sigma}(\bfa).
\ee

\subsection{Dualities of SQCD}

In this section we consider the index matching for the duality $\cG \leftrightarrow \cG_s$. This duality is obtained by Higgsing the flavor symmetry subgroup $SU(N)_A \times SU(N)_D$ to $U(1)_u \times U(1)_v$. In what follows, we will use the fugacity $u$ and $v$ for the flavor symmetries $U(1)_u$ and $U(1)_v$ respectively. The theory $\cG$ is the familiar $\cN{=}1$ $SU(N)$ gauge theory with $2N$ flavors. Its index is obtained using the prescription eq. \eqref{general-prescription} to close the punctures:
\be
{\cal I}_{\cG}(u,\bfb;\bfc,v)=\frac{K_\mn^{+}(u)K^{+}(\bfb)K^{-}(\bfc)K_\mn^{-}(v)}{K_{\varnothing}^{+}K_{\varnothing}^{-}}\sum_{\lambda}\frac{\Psi_{\lambda}^{\mn+}(u)\Psi_{\lambda}(\bfb)\Psi_{\lambda}(\bfc)\Psi_{\lambda}^{\mn-}(v)}{\Psi_{\lambda}^{\varnothing+}\Psi_{\lambda}^{\varnothing-}}.
\ee
The dual side is obtained by Higgsing the symmetries of ${\cal T}_s$. As explained in section \ref{newdual}, the resulting theory consists of  a charge-shifted  copy of ${\cal T}$, as given by \eqref{Usshift}, coupled to gauge singlet fields $M$. The index of ${\cal T}$ after this charge shift is ${\cal I}_{\cal T}(\bfc, v(\xi^{-1}\sqrt{pq})^\mn;u(\xi\sqrt{pq})^\mn, \bfb)$. The gauge singlet fields $M$ contain $M_B$ and $M_C$  which contribute $M^+(\bfb)M^-(\bfc)$ to the index as before. In addition, there are also fields $(M_{A})_{j,-j,k}$ and $(M_{D})_{j,-j,k}$ resulting from the Higgsing at punctures $A$ and $D$. More generally, such fields coupled to a generic puncture of type $\Lambda$ contribute:
\be
M_{\Lambda}^\sigma(\bfu)=\prod_{j}\mbox{PE}\left(\frac{(\xi^\sigma\sqrt{pq})^{1+j}-pq/(\xi^\sigma\sqrt{pq})^{1+j}}{(1-p)(1-q)}\chi_{R_{j}}(\bf u)\right).
\ee
With this at hand, we write the index of $\cG_s$ as:
\be
{\cal I}_{\cG_s}=M_\mn^{+}(u)M^{+}(\bfb)M^{-}(\bfc)M_\mn^{-}(v){\cal I}_{\cal T}(\bfc, v(\xi^{-1}\sqrt{pq})^\mn;u(\xi\sqrt{pq})^\mn, \bfb).
\ee
Indeed we see that ${\cal I}_\cG={\cal I}_{\cG_s}$, thanks to the generalization of the identity eq. \eqref{cid},
\be\label{cool-identity}
M_{\Lambda}^{\sigma}(\bfu)K^{-\sigma}(\bfu(\xi^\sigma\sqrt{pq})^\Lambda)=K_{\Lambda}^{\sigma}(\bfu).
\ee
for any partition $\Lambda$. We get the desired result after substituting $\Lambda=\mn$. 

We can also consider more general type of Higgsing on both sides of the duality ${\cal T}\leftrightarrow {\cal T}_s$. The general identity eq. \eqref{cool-identity} enables us to demonstrate the index equality for all resulting dualities.

%%%%%%%%%%%%%%%% section 5 %%%%%%%%%%%%%%%%%%%%%%%%%%%%%%%%%%%%%%%%%%%%%
\section{Quivers}
\label{sec:quiver}
In this section, we consider a generalization of the dualities studied in the previous sections to quiver gauge theories constructed from the $T_{N}$ theories by gauging their flavor symmetries. Such quiver  theories without any flavor symmetry (except $U(1)_\CF$) were recently studied in \cite{Bah:2011vv,Bah:2012dg} and were shown to coincide with a subclass of $\cN{=}1$ SCFTs obtained from M5-brane compactification on Riemann surfaces \emph{without} punctures. There, the authors identified different quiver descriptions  which are expected to flow to the same low energy theory. The proposed IR equivalence of different UV descriptions can now simply be understood through a sequence of  more fundamental ${\cal T}\leftrightarrow{\cal T}_c$ and ${\cal T}\leftrightarrow{\cal T}_s$ dualities acting on individual nodes. We also consider quiver theories with flavor symmetries. We expect they  correspond to new $\cN{=}1$ fixed points obtained after compactifying M5-branes on Riemann surfaces \emph{with} punctures. The index of such theories is computed by the TQFT given in \cite{Beem:2012yn}.

%%%%%%%%%%%%%%%%%%%
\subsection{Without flavor symmetry}

The $T_N$ theory can be thought of as an $\cN{=}1$ SCFT with a $U(1)$ flavor symmetry. For every theory $T_N^{(i)}$ in the quiver, let us call this flavor symmetry $J^{(i)}$. It is given in terms of its $\cN{=}2$ charges as $J^{(i)}=R_{\cN{=}2}^{(i)}-2I_3^{(i)}$. When theories $T_N^{(1)}$ and $T_N^{(2)}$ are coupled using the $\cN{=}1$ vector multiplet, only the off-diagonal combination of $J^{(1)}$ and $J^{(2)}$ survives in the resulting SCFT. On the other hand, their coupling with $\cN{=}2$ vector multiplet preserves the diagonal combination. The $J$ symmetry of the chiral field inside the $\cN{=}2$ vector multiplet also contributes to the conserved symmetry with the same sign as $T_N$. In order to consider more general quivers, we best label each $T_N^{(i)}$ theory with a sign $\sigma^{(i)}$. This is the sign with which $J^{(i)}$ contributes to the conserved symmetry $\CF$. Let us also define the sign of $\cN{=}2$ vector multiplet as the sign of the $\CF$ charge of the chiral field. The above discussion then gives us the general rules to construct an $\cN{=}1$ SCFTs with $T_N$ theories:
\begin{itemize}
\item Couple $T_N$ theories of opposite signs with $\cN{=}1$ vector multiplet
\item Couple $T_N$ theories of same sign $\sigma$ with $\cN{=}2$ vector multiplet of sign $\sigma$.
\end{itemize}
These rules were also obtained in \cite{Bah:2011vv,Bah:2012dg}. 

\begin{figure}[h]
\centering
\begin{tikzpicture}[scale=.4,baseline=(current bounding box.east)]\footnotesize
\draw (0,0)--(-1,-1)--(-1,1) node[pos=.5,left]{$T_N^{(1)}$} --cycle;
\draw (-0.6,0) node {$+$};
\draw (-2,2) node[anchor=east] {$A$} --(-1,1);
\draw[dotted] (-4,4)--(-2,2);
\draw (-2,-2) node[anchor=east] {$B$} --(-1,-1);
\draw[dotted] (-4,-4)--(-2,-2);
\draw (0,0)--(3,0) node[pos=.5,above]{$W$};
\draw (3,0)--(4,-1)--(5,0)--(3,0) node[pos=.5,above]{$T_N^{(2)}$};
\draw (4,-1)--(4,-3) node[right]{$C$};
\draw[dotted] (4,-3)--(4,-5);
\draw (4,-0.3) node {$-$};
\draw (5,0)--(8,0) node[pos=.5,above]{$Z$};
\draw (8,0)--(9,-1)--(9,1) node[pos=.5,right]{$T_N^{(3)}$} --cycle;
\draw (8.6,0) node {$+$};
\draw (10,2) node[anchor=west] {$D$} --(9,1);
\draw[dotted] (12,4)--(10,2);
\draw (10,-2) node[anchor=west] {$E$} --(9,-1);
\draw[dotted] (12,-4)--(10,-2);
\end{tikzpicture}
\caption{A piece of the quiver.}
\label{GenDual01}
\end{figure}
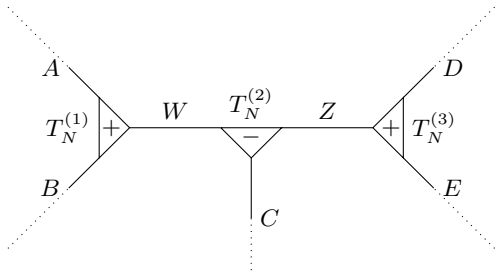
\begin{figure}[h]
\centering
\begin{tikzpicture}[scale=.4,baseline=(current bounding box.east)]\footnotesize
\draw (0,0)--(-1,-1)--(-1,1) node[pos=.5,left]{$T_N^{(1)}$} --cycle;
\draw (-0.6,0) node {$-$};
\draw (-2,2) node[anchor=east] {$M_A-A$} --(-1,1);
\draw[dotted] (-4,4)--(-2,2);
\draw (-2,-2) node[anchor=east] {$M_B-B$} --(-1,-1);
\draw[dotted] (-4,-4)--(-2,-2);
\draw (0,0)--(3,0) node[pos=.5,above]{$W$};
\draw (3,0)--(4,-1)--(5,0)--(3,0) node[pos=.5,above]{$T_N^{(2)}$};
\draw (4,-1)--(4,-3) node[right]{$C-M_C$};
\draw[dotted] (4,-3)--(4,-5);
\draw (4,-0.3) node {$+$};
\draw (5,0)--(8,0) node[pos=.5,above]{$M_Z{-}Z$};
\draw (8,0)--(9,-1)--(9,1) node[pos=.5,right]{$T_N^{(3)}$} --cycle;
\draw (8.6,0) node {$+$};
\draw (10,2) node[anchor=west] {$D$} --(9,1);
\draw[dotted] (12,4)--(10,2);
\draw (10,-2) node[anchor=west] {$E$} --(9,-1);
\draw[dotted] (12,-4)--(10,-2);
\end{tikzpicture}
\caption{The quiver after the swap duality at node $W$.}
\label{GenDual02}
\end{figure}
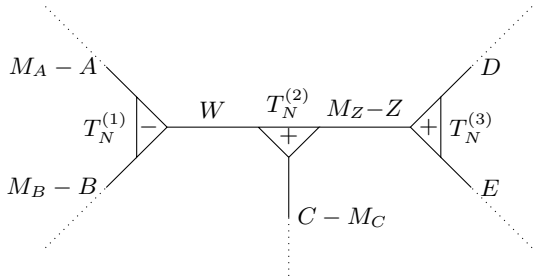

Now let us consider a general quiver of $T_N$ theories without flavor symmetry that obeys the above rules. In figure \ref{GenDual01} we have zoomed in on a random piece of this quiver. The theories $T_N^{(1)}$ and $T_N^{(2)}$ have opposite signs and hence are coupled at node $W$ by an $\cN{=}1$ vector multiplet. Similarly, at node $Z$, the theories $T_N^{(2)}$ and $T_N^{(3)}$ are coupled with an $\cN{=}1$ vector multiplet. The coupling involves the superpotential 
\be\label{quiverW}
W=\tr \mu^{(1)}_W \mu^{(2)}_W+\tr \mu^{(2)}_Z \mu^{(3)}_Z.
\ee
Let us apply ${\cal T}\leftrightarrow{\cal T}_s$ at node $W$. This duality \emph{swaps} the signs of the $T_N^{(1)}$ and $T_N^{(2)}$ theories. The operators $\mu^{(1)}_{X}$ ($X=A,B$) and $\mu^{(2)}_{X}$ ($X=C,Z$) transforming in the adjoint of $SU(N)_X$ are mapped to new chiral fields $M_X$. The new quiver is shown in figure \ref{GenDual02}.
Let us focus on the field $M_Z$. It is coupled to the dual ${\hat \mu}^{(2)}_Z$ operators through the superpotential $\tr {\hat \mu}^{(2)}_Z M_Z$. From the equation \eqref{quiverW}, we also see that it gets coupled to $\mu^{(3)}_Z$ through the superpotential $\tr \mu^{(3)}_Z M_Z$. The superpotential after the duality is
\be\label{quivernewW}
W=\tr \mu^{(1)}_W \mu^{(2)}_W+\tr  M_Z ({\hat \mu}^{(2)}_Z+\mu^{(3)}_Z).
\ee
The second term has the form of the characteristic superpotential coupling of the adjoint chiral field of $\cN{=}2$ vector multiplet at node $Z$. This shows that the field $M_Z$ actually combines with the $\cN{=}1$ vector multiplet at node $Z$ to form an $\cN{=}2$ vector multiplet. This is fortunate because $T_N^{(2)}$ and $T_N^{(3)}$  have the same sign after the duality and hence require an $\cN{=}2$ vector multiplet coupling. The new quiver satisfies the SCFT rules and describes the same low energy physics. We have derived a different but equivalent quiver description where the $+$ and $-$ signs of the adjacent $T_N$ blocks are swapped. Note that the total number of $+$ signs and total number of $-$ signs has remained the same. At the $\cN{=}2$ node, we can use S-duality to further rearrange the edges of the quiver.

\begin{figure}[h]
\centering
\begin{tikzpicture}[scale=.4,baseline=(current bounding box.east)]\footnotesize
\draw (1,2) node[anchor=east] {$A$} --(3,0);
\draw[dotted] (-1,4)--(1,2);
\draw (3,0)--(4,-1)--(5,0)--(3,0) node[pos=.5,above]{$T_N^{(2)}$};
\draw (4,-0.3) node {$+$};
\draw (4,-1)--(4,-3) node[pos=.5,left]{$W$};
\draw (4,-3)--(3,-4)--(5,-4) node[pos=.5,below]{$T_N^{(1)}$} --cycle;
\draw (4,-3.6) node {$-$};
\draw (2,-5) node[anchor=east] {$M_B-B$} --(3,-4);
\draw[dotted] (0,-7)--(2,-5);
\draw (6,-5) node[anchor=west] {$C$} --(5,-4);
\draw[dotted] (8,-7)--(6,-5);
\draw (5,0)--(8,0) node[pos=.5,above]{$M_Z{-}Z$};
\draw (8,0)--(9,-1)--(9,1) node[pos=.5,right]{$T_N^{(3)}$} --cycle;
\draw (8.6,0) node {$+$};
\draw (10,2) node[anchor=west] {$D$} --(9,1);
\draw[dotted] (12,4)--(10,2);
\draw (10,-2) node[anchor=west] {$E$} --(9,-1);
\draw[dotted] (12,-4)--(10,-2);
\end{tikzpicture}
\caption{The quiver after the crossing duality at node $W$.}
\label{GenDual03}
\end{figure}
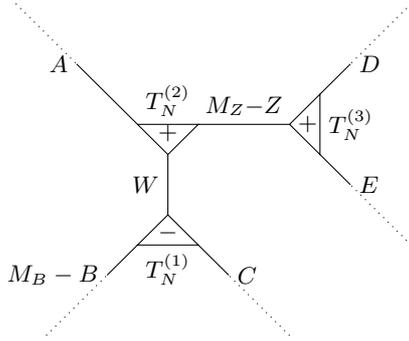

We can also apply the duality ${\cal T}\leftrightarrow{\cal T}_c$ or ${\cal T}\leftrightarrow{\cal T}_{c^\prime}$ at node $W$ in figure \ref{GenDual01}. The $M$ fields resulting from the crossing duality again combine with the $\cN{=}1$ vector multiplet at node $Z$ to form the $\cN{=}2$ vector multiplet. This also results in a quiver that satisfies the above rules and hence flows to an equivalent SCFT, see figure \ref{GenDual03}. 
These fundamental dualities at individual nodes are sufficient to reshuffle the $+$ and $-$ signs in all possible ways keeping the total number each constant. This can be considered a derivation of the IR equivalence of different quiver theories proposed in \cite{Bah:2011vv,Bah:2012dg}. In the description of these $\cN{=}1$ SCFTs as M5-brane compactifications, the numbers of $+$ and $-$ signs in the quiver are the degrees $p$ and $q$ of the two line bundles normal to the Riemann surface.

%%%%%%%%%%%%%%%%%%%%%%%%%%
\subsection{With flavor symmetry}

Realizing the IR equivalence of different quiver description using local dualities has a crucial advantage: it allows to consider  dualities of theories with flavor symmetries. To classify such quivers, in addition to attaching a sign to component  $T_N$ theories, we also need to attach a sign to each $SU(N)$ flavor symmetry. As we expect such quivers to correspond to M5-brane compactifications on Riemann surfaces with punctures, attaching a sign to $SU(N)$ flavor symmetry corresponds to attaching a sign to a puncture. We claim that the data: $(p,q,n^+,n^-)$ uniquely classifies the inequivalent low energy theories. Here, $p$ ($q$) is the number of $T_N$ theories with $+$($-$) signs and $n^+$ ($n^-$) is the number of punctures with $+$($-$) signs. Let us construct a quiver which realizes this data. We give the following prescription to read off the UV description.
\begin{itemize}
\item When a puncture $X$ of type $\sigma$ is attached to a $T_N$ theory of type $\sigma$, it represents the ordinary $SU(N)_X$ flavor symmetry of the $T_N$ theory. 
\item When a puncture $X$ of type $\sigma$ is attached to a $T_N$ theory of type $-\sigma$, we have additional chiral fields $M_X$ that transform in the adjoint representation of $SU(N)_X$. They are coupled to the $\mu_X$ operators of the $T_N$ theory through a superpotential coupling $\tr \mu_X M_X$.
\end{itemize}
We conjecture that all UV theories with the same $(p,q,n^+,n^-)$ data, flow to the same fixed point. We can further Higgs the $SU(N)$ flavor symmetries in each description with vevs corresponding to arbitrary partitions. This results in a significant extension of the duality web. The matching of the superconformal index for the duality $\cG\leftrightarrow \cG_s$  in section \ref{sec:index} can be straightforwardly generalized to show the index matching for all conjectured dualities resulting from such Higgsing. The $\cN{=}1$ generalized  quiver gauge theories with $N=2$ were first considered in \cite{Maruyoshi:2009uk}.
 
%%%%%%%%%%%%%%%%%%%%%%%%%%%%%%%%%%%%%%%%%%%
\subsection{Central charges}

The $U(1)_R$ and $U(1)_\CF$ symmetries mix to define the new superconformal R symmetry in the infrared,
\bea
R(\epsilon)
=     R + \epsilon \CF.
\eea
The coefficient $\epsilon$ is determined by maximizing the trial $a$ function \cite{Intriligator:2003jj}
\bea
a(\epsilon)
=     \frac{3}{32} (3 \Tr R(\epsilon)^{3} - \Tr R(\epsilon)).
\eea
Here we calculate this function of the $\CN{=}1$ quiver gauge theory constructed in the previous subsection. For simplicity, we choose all the punctures to be maximal.

Let $a_{T_{N}}^\sigma$ be the contribution of the $T_{N}$ theory with sign $\sigma$ to $a(\epsilon)$. Also let $a_{V}$ and $a_{\chi}^{\sigma}$ be the contributions of the $\CN{=}1$ vector multiplet and the $\CN{=}1$ adjoint chiral multiplet with $\U(1)_{\CF}$ charge $2\sigma$ respectively. As in \cite{Bah:2012dg}, they are given by
\bea
a_{T_{N}}^\sigma(\epsilon)
&=&    \frac{3}{32} A(2\sigma\epsilon),
\nonumber \\
A(\epsilon)
&=&    \left( \frac{3}{8} (1 + \epsilon)^{3} - \frac{1}{2} (1 + \epsilon) \right) \Tr R^{3}_{\CN{=}2}
+ \frac{9}{2} (1 + \epsilon)(1 - \epsilon)^{2} \Tr R_{\CN{=}2}I_{3}^{2},
\nonumber \\
a_{V}
&=&    \frac{6}{32} (N^{2} - 1), ~~~~~~
a_{\chi}^\sigma
=     \frac{3}{32} (N^{2} - 1) \left( 24 \epsilon^{3} - 2\epsilon \right) \sigma.
\eea
In order to calculate $a(\epsilon)$, we need to know the number of $T_N$ theories, the number of vector multiplets and the number of adjoint chiral fields. The counting of the first two is straightforward. The adjoint chiral fields come either as $M$ fields associated to flavor symmetries (punctures) or as the chiral fields in the $\cN{=}2$ vector multiplet. Let $n_\sigma^{\sigma^\prime}$ be the number of punctures of sign $\sigma$ attached to the $T_N$ theory of sign $\sigma^\prime$. As discussed in the previous subsection, the number of $M$ fields with $\CF$ charges $+2$ and $-2$ is $n_-^+$ and $n_+^-$ respectively,
\bea
a_{M}
=     \frac{3}{32}(N^{2}-1) \left( 24 \epsilon^{3} - 2\epsilon \right)
\left( n_{-}^{+} - n_{+}^{-} \right).
\eea
The contribution of the gauge adjoint multiplets $\phi$ in $\CN{=}2$ vector multiplets can be obtained as follows: The number of $T_N$ theories with sign $+$ and $-$ is $p$ and $q$ respectively. The number of punctures of $T_N$ theory with $+$ ($-$) sign that have been gauged is $N_+:=3p-n_-^+-n_+^+$($N_-:=3p-n_-^--n_+^-$). The adjoint chiral fields $\phi$ with opposite $\CF$ charge contribute oppositely to $a(\epsilon)$. The net contribution to $a(\epsilon)$ only comes from $N_+-N_-$ chiral fields.
\bea
a_\phi
=     \frac{3}{32}(N^{2}-1) \left( 24 \epsilon^{3} - 2\epsilon \right) \frac{N_{+} - N_{-}}{2}.
\eea
By summing up all the contributions, we get
\bea
a(\epsilon)
&=&    \frac{3}{32} \left( p A(2\epsilon) + q A(-2\epsilon) \right)
+ \frac{6}{32} (N^{2} - 1) (3g - 3 + n_++n_-)
\nonumber \\
& &  + \frac{3}{64}(N^{2}-1) \left( 24 \epsilon^{3} - 2\epsilon \right)
\left( 3p - 3q - n_{+} + n_{-} \right).
\eea
Note that the final result depends only on $p$, $q$, $n_{+}$ and $n_{-}$. For $n_+=n_-=0$, this expression reduces to the expression for $a(\epsilon)$ obtained in \cite{Bah:2012dg}. The IR R charge can be obtained by maximizing this function. Thus for the theories with the same $p$, $q$, $n_{+}$ and $n_{-}$, the IR R-symmetry is the same.

%%%%%%%%%%%%%%%%%%%%%%%%%%%%%%%%%%%%  acknowledgements
\section*{Acknowledgments}
It is a pleasure to thank Chris Beem, Yu Nakayama, Shlomo Razamat for useful comments and discussions. K.M. and Y.T. would also like to thank the hospitality of Mathematical Physics group in Osaka City University where part of this work was carried out. The work of A.G. is supported in part by the John A. McCone fellowship and by DOE grant DE-FG02-92-ER40701. The work of K.M. is supported by JSPS postdoctoral fellowships for research abroad. The work of Y.T. is supported in part by World Premier International Research Center Initiative  (WPI Initiative),  MEXT, Japan through the Institute for the Physics and Mathematics of the Universe, the University of Tokyo. The work of W.Y. is supported in part by the Sherman Fairchild scholarship and by DOE grant DE-FG02-92-ER40701.

\bibliographystyle{ytphys}
%\small\baselineskip=.9\baselineskip
\bibliography{ref}

\end{document}